\documentstyle[useAMS,graphicx,longtable]{mn2e}

\newcommand{\hMpc}{{\ifmmode{h^{-1}{\rm Mpc}}\else{$h^{-1}$Mpc}\fi}}
\newcommand{\hkpc}{{\ifmmode{h^{-1}{\rm kpc}}\else{$h^{-1}$kpc}\fi}}

\def\approxlt{\mathrel{\spose{\lower 3pt\hbox{$\sim$}}
        \raise 2.0pt\hbox{$<$}}}
\def\approxgt{\mathrel{\spose{\lower 3pt\hbox{$\sim$}}
        \raise 2.0pt\hbox{$>$}}}
\def\approxpropto{\mathrel{\spose{\lower 3pt\hbox{$\sim$}}
        \raise 2.0pt\hbox{$\propto$}}}

\title{Multiperiodic Galactic field RR Lyrae stars in the ASAS catalog}

\author[D. M. Szczygie{\l{}} and D. C. Fabrycky]
{D. M. Szczygie{\l{}}$^{1}$\thanks{e-mail: dszczyg@astrouw.edu.pl}
and D. C. Fabrycky$^{2}$\thanks{e-mail: dfab@astro.princeton.edu}\\
$^{1}$Warsaw University Observatory, Al.Ujazdowskie 4, 00-478 Warsaw, Poland\\
$^{2}$Princeton University Observatory, Peyton Hall, Princeton, NJ 08544, USA}

\date{Accepted --.
      Received -- ;
      in original form --}

\pubyear{2007}

\begin{document}

\maketitle

\label{firstpage}

\begin{abstract}

The All Sky Automated Survey (ASAS) monitors bright stars (8 mag $< V <$ 14 mag)
south of declination ${\rm + 28^{\circ}}$.  The ASAS Catalogue of Variable Stars (ACVS) 
presently contains 50,099 objects; among them are 2212 objects classified as RR Lyrae 
pulsating variables. We use ASAS photometric $V$ band data to search for 
multiperiodicity in those stars. We find that 73 of 1435 RRab stars and 49 of 756 RRc stars
exhibit the Blazhko effect. 
We observe a deficiency of RRab Blazhko variables with main pulsation periods greater than
0.65 days.
The Blazhko periods of RRc stars exhibit a strongly bimodal distribution.
During our study we discovered the Blazhko effect with multiple periods in object ASAS 
050747-3351.9 = SU Col. Blazhko periods of  $89.3$ d and $65.8$ d and a candidate of $29.5$ d 
were identified with periodogram peaks near the first three harmonics of the main pulsation.
These observations may inspire new models of the Blazhko effect, which has eluded a consistent 
theory since its discovery about one hundred years ago.
Long term light curve changes were found in 29 stars.  We also found 19 Galactic double mode pulsators (RRd),
of which 4 are new discoveries, raising the number of ASAS discoveries of such objects to 16,
out of 27 known in the field of our Galaxy.

\end{abstract}


\section{Introduction}
\label{sect:intro}

The discovery of RR Lyrae stars in star clusters (Bailey 1902) led to some of the first studies 
of radial pulsations in stars (Shapley 1914, Eddington 1917) as well as distance determinations 
based on stellar variability (Shapley 1918).  Recent surveys have searched for 
substructure in the Milky Way, attributed to the incorporation of satellite galaxies, via 
overdensities of RR Lyrae stars (Duffau et al. 2006).  
Their pulsation means they lie in the instability strip of the Hertzprung-Russell diagram, 
the extent (and metallicity dependence) of which may be evaluated based on a statistical sample
of such stars.  Finally, RR Lyrae stars are found to pulsate not just with a single period, but
in multiple radial and non-radial modes of different period, which gives constraints both to
pulsation theory and to internal structure calculations. 
The current study presents the
photometric signal of these multiperiodic phenomena in a large sample of RR Lyrae stars.

In Section \ref{sec:asas} we shortly describe ASAS experiment and its database, as well as
the RR Lyrae sample derived from the ASAS Catalogue of Variable Stars.
Sections \ref{sec:be} and \ref{sec:rrd} contain the method and results of multiperiodicity search, 
together with the discussion.
Finally in Section \ref{sec:sum} we summarize the results.

\section{ASAS data}
\label{sec:asas}

\subsection{The All Sky Automated Survey}

The ASAS project is located in Las Campanas Observatory in Chile and currently consists
of two small, wide-field telescopes (200/2.8) with ${\rm 2K \times 2K }$ CCD cameras with 
${\rm 15 \mu }$ pixel size from Apogee, observing in standard \emph{V} and \emph{I} filters.
The main goal of the ASAS project is to monitor the whole available bright (8 mag $< V <$ 14 mag)
sky and search for variability. It has been observing south of 
declination $+ 28^{\circ}$ (almost 75\% of the whole sky) since 2000, and a smaller part 
since 1997, covering the available sky every 1-3 nights.  The project is ongoing, but has 
already yielded a few hundred photometric measurements per star.

Analysis of the \emph{V} band data taken over the last 6 years resulted in the ASAS 
Catalog of Variable Stars (ACVS), which presently contains 50,099 variable stars. The 
\emph{I} band data is currently being processed and is not yet included in the catalog. ACVS 
is available both as an online database and a downloadable text file. For details on ASAS 
equipment and catalogues see Pojma\'nski (1997, 1998, 2000, 2002, 2003), Pojma\'nski and 
Maciejewski (2004, 2005), Pojma\'nski, Pilecki and Szczygie{\l{}} (2005). All the ASAS data that is 
public domain, as well as an online catalogue, are available on the WWW:
\centerline{ http://www.astrouw.edu.pl/\~{}gp/asas/asas.html }
\centerline{ http://archive.princeton.edu/\~{}asas/ }

\subsection{RR Lyrae Sample in ACVS}

\begin{figure}
\vspace{-7.5cm}
\includegraphics[scale=.45]{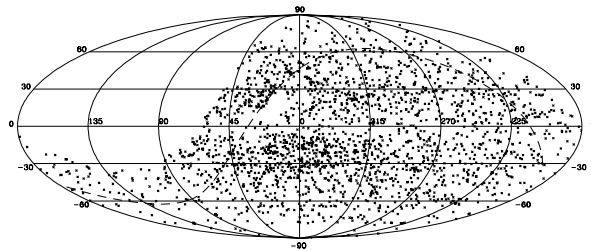}
\vspace{-0.5cm}
\caption{Distribution of RR Lyrae stars from ASAS Catalogue of Variable Stars (ACVS) in galactic coordinates.
The dashed line is the equatorial plane. \label{sphere}}
\end{figure}

ACVS contains a sample of 2212 RR Lyrae stars: 1455 of RRab 
type and 757 of RRc type. In variable stars with ambiguous classification, we used those 
for which RR Lyr type is the most probable. Their distribution across the sky in galactic 
coordinates is shown in Figure \ref{sphere}.

The classification procedure of variable stars in ASAS was based mainly on Fourier 
parameters of the light curve, JHK colours from 2MASS, and the object's period as described 
in Pojma\'nski (2002). The full catalogue consists of five parts created between 2002 and 2005, 
so the classification of the light curves as RR Lyrae stars was based on measurements taken 
from different time baselines. But we used the longest available baseline (for most objects 
between HJD 2451800 to HJD 2453800) to refine their periods, which was especially useful for
objects from the oldest catalogues. We also used all available data points of reasonable quality
(i.e. with flags A and B) in the search for multiperiodic behaviour.  

Because the ASAS classification was automatic and allows multiple type assignment we decided 
to roughly inspect the light curves visually to remove variable stars of other types, 
which were mistaken for RR Lyrae stars. That left us with 1435 RRab and 756 RRc stars. 
It is important to mention that the inspection was very superficial and excluded 
only obvious misclassifications among RRab objects, e.g. MIRA type variables. 
The light curves of RRc type are very similar to EC (i.e. eclipsing contact binary) type, 
we simply trust the automatic classification, accepting a high rate of misidentification.

\begin{figure}
\begin{center}
\includegraphics[scale=.35]{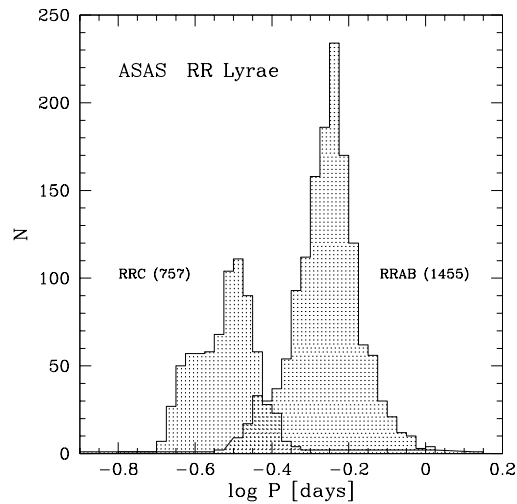}
\end{center}
\caption{The distribution of periods of the RR Lyrae sample. \label{periods}}
\end{figure}

\begin{figure}
\begin{center}
\begin{tabular}{c c}
\includegraphics[scale=.19]{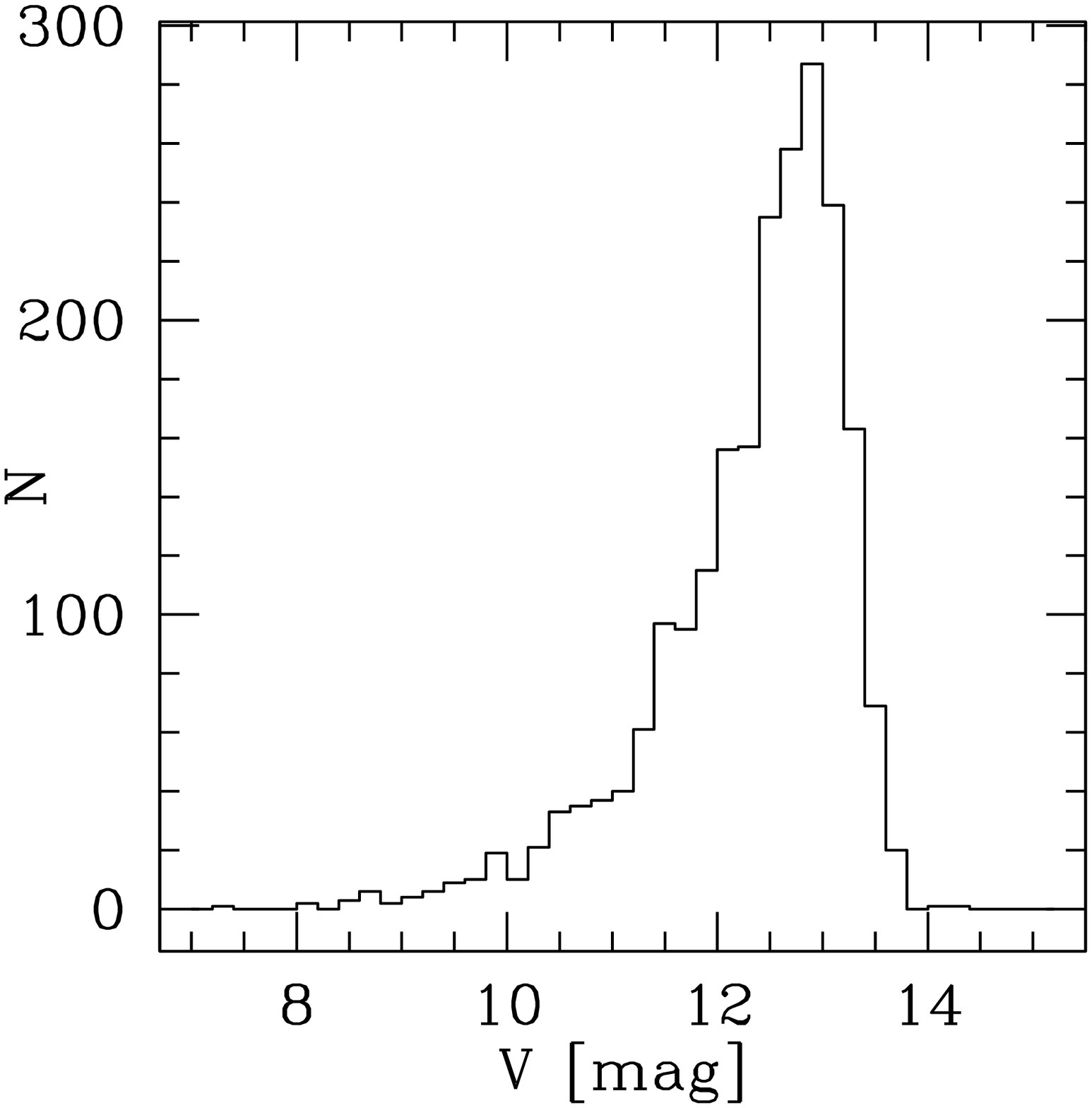} &
\includegraphics[scale=.19]{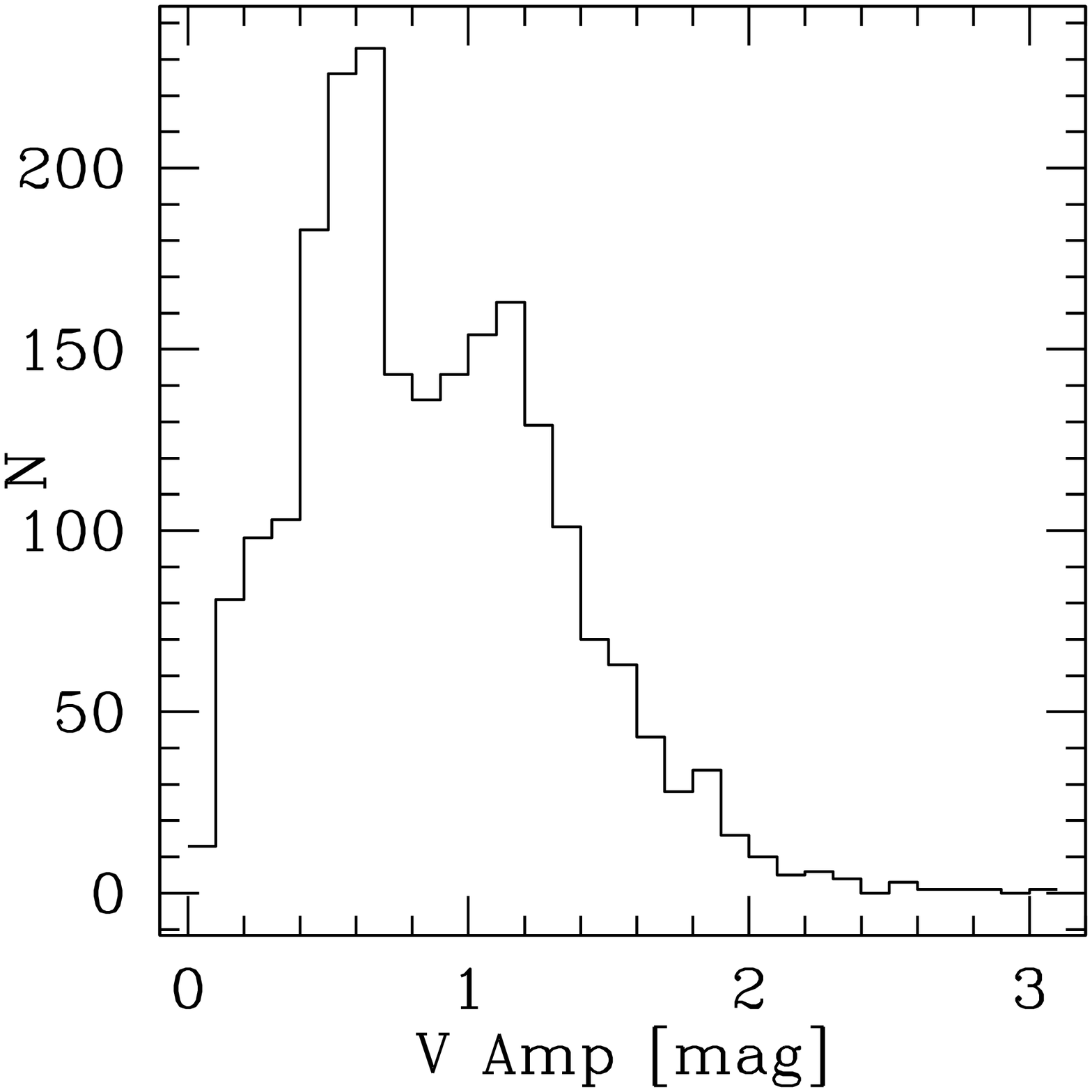} \\
\end{tabular}
\end{center}
\caption{The distribution of $V$ magnitudes in maximum light and $V$ amplitudes of the RR Lyrae sample. \label{mag.amp}}
\end{figure}

The distribution of periods of our RR Lyrae sample is shown in Figure \ref{periods}. Figure 
\ref{mag.amp} presents two histograms also based on the sample, one of magnitudes in maximum 
light and the other one of the amplitudes of the light curves.


\section{Blazhko effect}
\label{sec:be}

The shape of the light curves of many RR Lyrae stars changes in a nearly
periodic fashion, with time-scales of between about five and a thousand or more days; 
this is called the Blazhko effect (Blazhko 1907, Smith 1995).
Since this shape change is equivalent to periodic phase and amplitude changes of
the harmonics which make up the light curve, in the frequency domain it is identified
through sidebands (additional peaks) in the periodogram near the harmonics of the main pulsation.  
Physically these peaks may correspond to excited non-radial modes, but the physical
nature of the Blazhko effect is still unsettled a century after its discovery.
For some stars, which we label BL1, there is a single significant 
frequency near the main mode.  For other stars, which we label BL2, two peaks flank
the main frequency with approximately equal frequency spacing (Mizerski 2003, Alcock et al. 2003).  
There are also some stars with two additional peaks which do not form an equidistant triplet with 
the main pulsation and others with more than two peaks.

\subsection{Identifying extra frequencies}

The method we use for discovering Blazhko stars is a slightly modified version of the method
applied by Collinge et al. (2006) to Optical Gravitational Lensing Experiment data.
That paper gives a full introduction to the method which we discuss more concisely here.  

We first subtract a Fourier series of best-fitting harmonics of the main pulsation period
from the data, which is called prewhitening.  The number of harmonics to fit was chosen
by the method of Kovacs (2005) and depends on an empirical signal to noise level.
We next run the CLEAN algorithm (Roberts et al., 1987) on the prewhitened light curves,
using 200 iterations and a gain of 0.5, on the frequency 
band [0--6~day$^{-1}$] with a frequency resolution of $0.125/T$, where $T$ is the 
overall time baseline of the observations for the given light curve. The peaks are sorted by amplitude,
and the noise level is indicated by the 15th largest peak (${\cal{A}}_{15}$).  By a Monte-Carlo method,
we determine how much bigger than that noise level a signal must be to classify as a detection.
The Blazhko effect has a period of longer than $\sim 5$ days, so we define the ``Blazhko range''
as the frequencies within 0.2~day$^{-1}$ of, but further than $1/T$ from, the main pulsation frequency. 

BL1 stars are those with a single peak with amplitude ${\cal{A}} > 2.0 {\cal{A}}_{15}$ 
in the Blazhko range of the CLEANed spectrum.  BL2 stars have two frequencies within the Blazhko range,
on opposite sides of the main frequency, equidistant to it to within $3.0/T$, and both frequencies 
must have amplitudes ${\cal{A}} > 1.15 {\cal{A}}_{15}$.  We follow Alcock et al. (2003) in defining a
class of period change (PC) stars that have one or more peaks within $1/T$ 
of the main frequency with amplitude(s) ${\cal{A}} > 1.3 {\cal{A}}_{15}$. The frequency criterion for PC
stars means that even if the phenomenon is periodic, we have not yet observed a full cycle; we have only
detected a changing light curve shape.  We note that a long-term amplitude change can also give an extra 
frequency component near the main frequency, so the PC class is imprecisely named.

During its many iterations, CLEAN produces somewhat biased amplitudes, so after selecting
the significant frequencies we refine the fit parameters with a non-linear solver
(the Levenberg-Marquardt algorithm; Press et al. 1989); the resulting frequencies and amplitudes
are reported in Tables \ref{table.bl1}, \ref{table.bl2}, and \ref{table.pc}. 
The best-fitting model is subtracted from the light curve and the amplitude of 
the highest remaining peak within 0.2~day$^{-1}$ of the main frequency is also 
reported in Tables \ref{table.bl1}, \ref{table.bl2}, and \ref{table.pc}. We do not consider this peak significant; it is 
reported to quantify the level of the noise for completeness studies.

We determined the amplitude thresholds cited above by creating 3 false 
catalogues, each of them based on the real data sets for each star. In these catalogues 
the magnitude residuals from the pre-whitening step were 
randomly swapped among the observation times. We search for peaks in the false 
catalogues with the CLEAN algorithm and set the amplitude thresholds to allow an
acceptable false alarm rate such that about 2\% of the stars in each of the 
categories BL1, BL2, and PC are likely to be spurious.

This method allowed us to determine unambiguous threshold values in the BL2 and PC 
groups, but it partially failed for the BL1 category. It turned out that false 
catalogues very often contain a frequency in Blazhko range, whose amplitude is above 
the chosen noise level. This is probably a result of high noise in ASAS data. And 
while the probability to find BL2s in false catalogues is low as it requires two 
evenly spaced frequencies, it grows rapidly for BL1s, where only one frequency is 
needed to create a false alarm.

So for BL1s we used a more conservative method of threshold determination. 
Each prewhitened light curve was separated into even-numbered 
and odd-numbered points in the chronologically ordered data set, making two light curves.
Then the search 
for Blazhko frequency was performed in all three light curves---the original and 
subdivided ones.  A frequency must appear at least in two of the three light curves 
to qualify as a true detection. This search 
was performed for each star in the real and the 3 false catalogues, which allowed us to 
determine an unambiguous threshold value, because in random realization of the 
light curve such coincidence is less likely to appear.

It turned out that in a real catalog 22 BL1 stars had a Blazhko frequency found in 
all 3 light curve realizations and 30 BL1 stars had it in two, while in false 
catalogues it never occurred in all 3. Those stars which had a BL1 frequency in all 3 
light curves are marked with a ``*'' in the last column of Table \ref{table.bl1}.

\subsection{Objects exhibiting Blazhko effect}
\label{sec:OeBe}

The number of unique stars in each of our categories is (BL1, BL2, PC) = (52, 70, 29); 
see Table \ref{table.bl_no} for details. 

\begin{table}
\caption{Numbers in different categories of RR Lyrae stars exhibiting Blazhko-type behaviour. 
See section \ref{sec:OeBe} for details \label{table.bl_no}}
\begin{center}
\begin{tabular}{l|c|c|c|c||r}
      & BL1 & BL2 & BL2x2 & PC & Total\\
\hline
RRab  &  32 &  41 &   1   & 14 & 1435 \\
RRc   &  20 &  29 &   1   & 15 &  756 \\
\hline
Total &  52 &  70 &   2   & 29 & 2191 \\
\end{tabular}
\end{center}
\end{table}

\begin{figure*}
\begin{center}
\begin{tabular}{c c c}
\includegraphics[scale=.19]{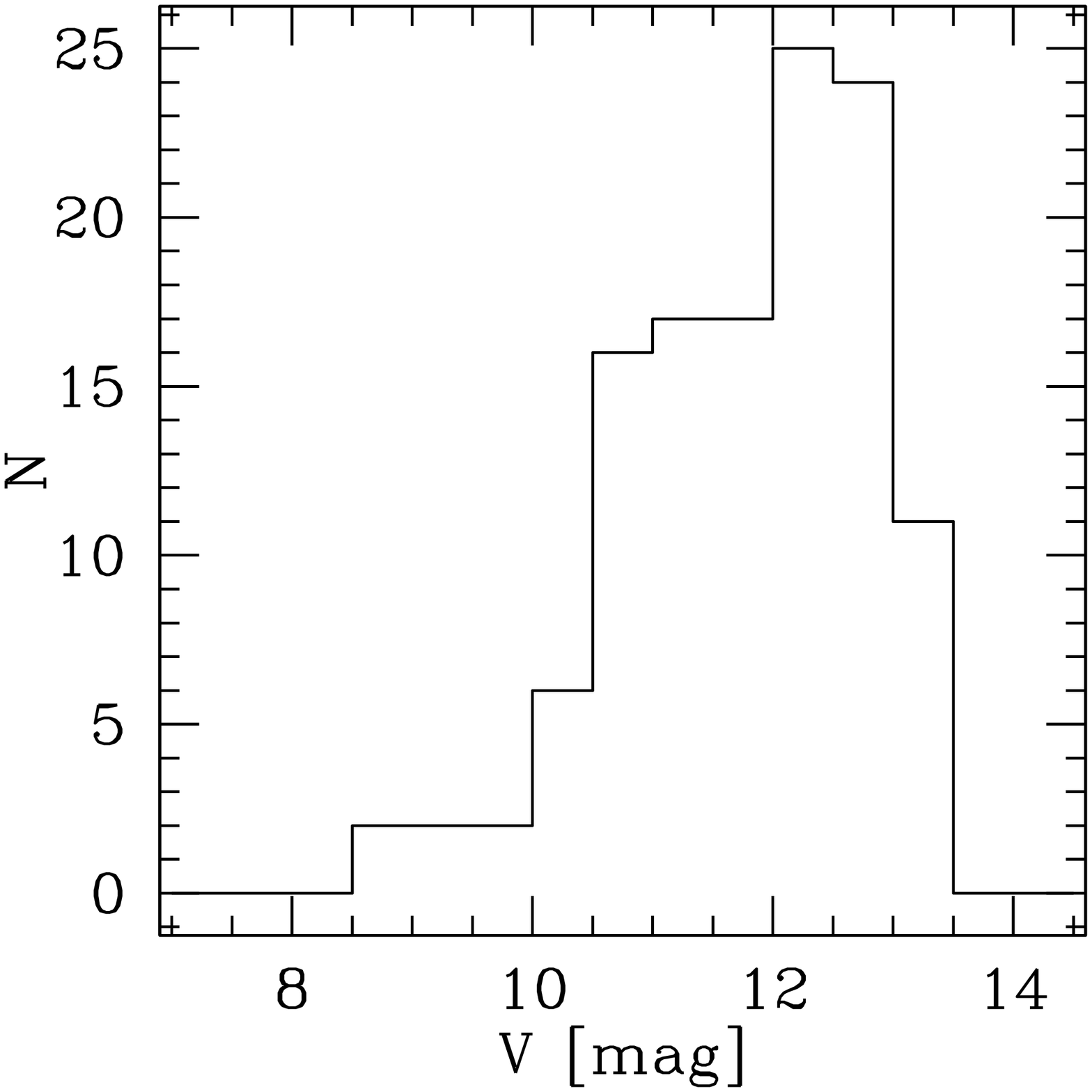} &
\includegraphics[scale=.19]{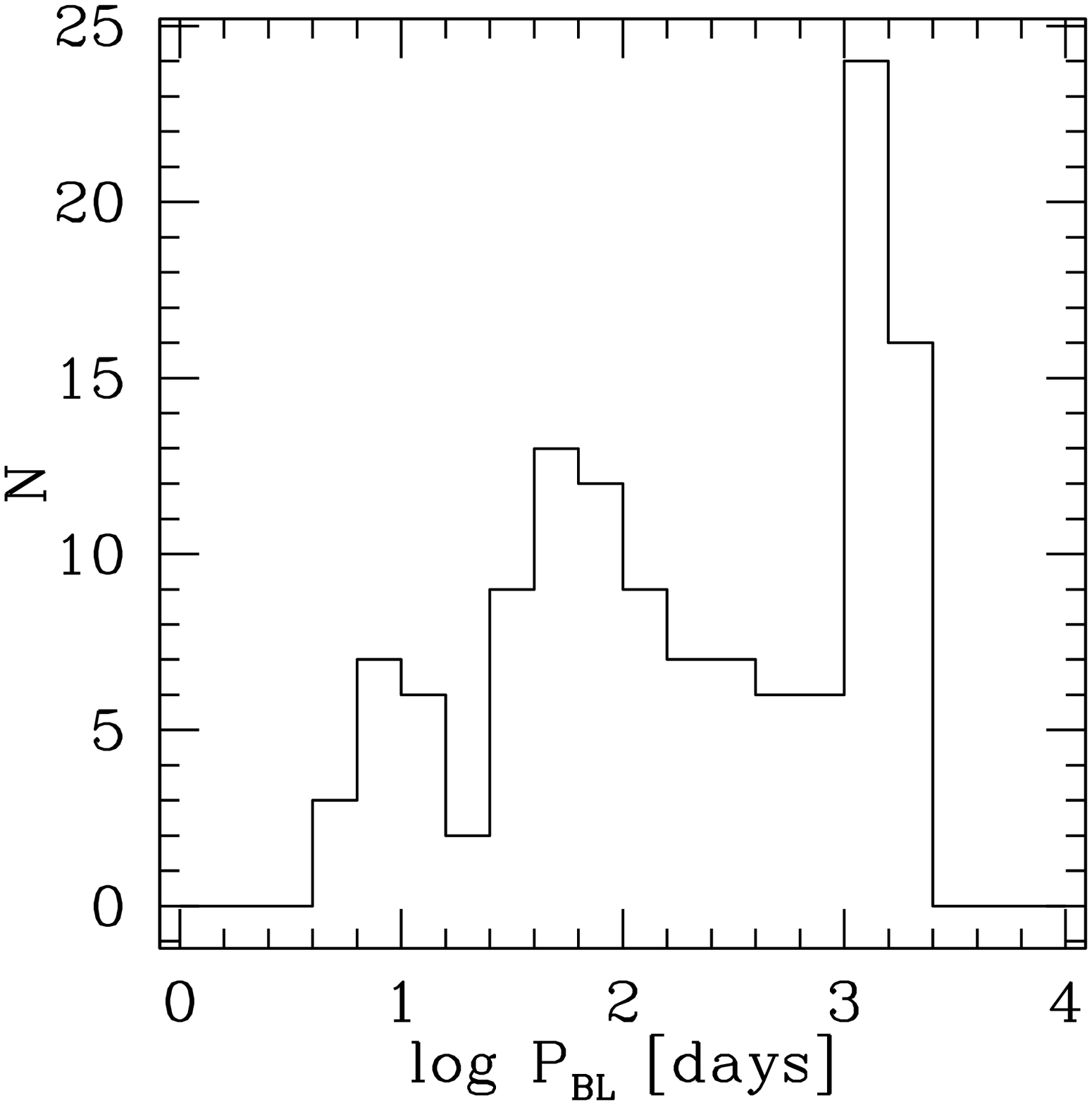} &
\includegraphics[scale=.19]{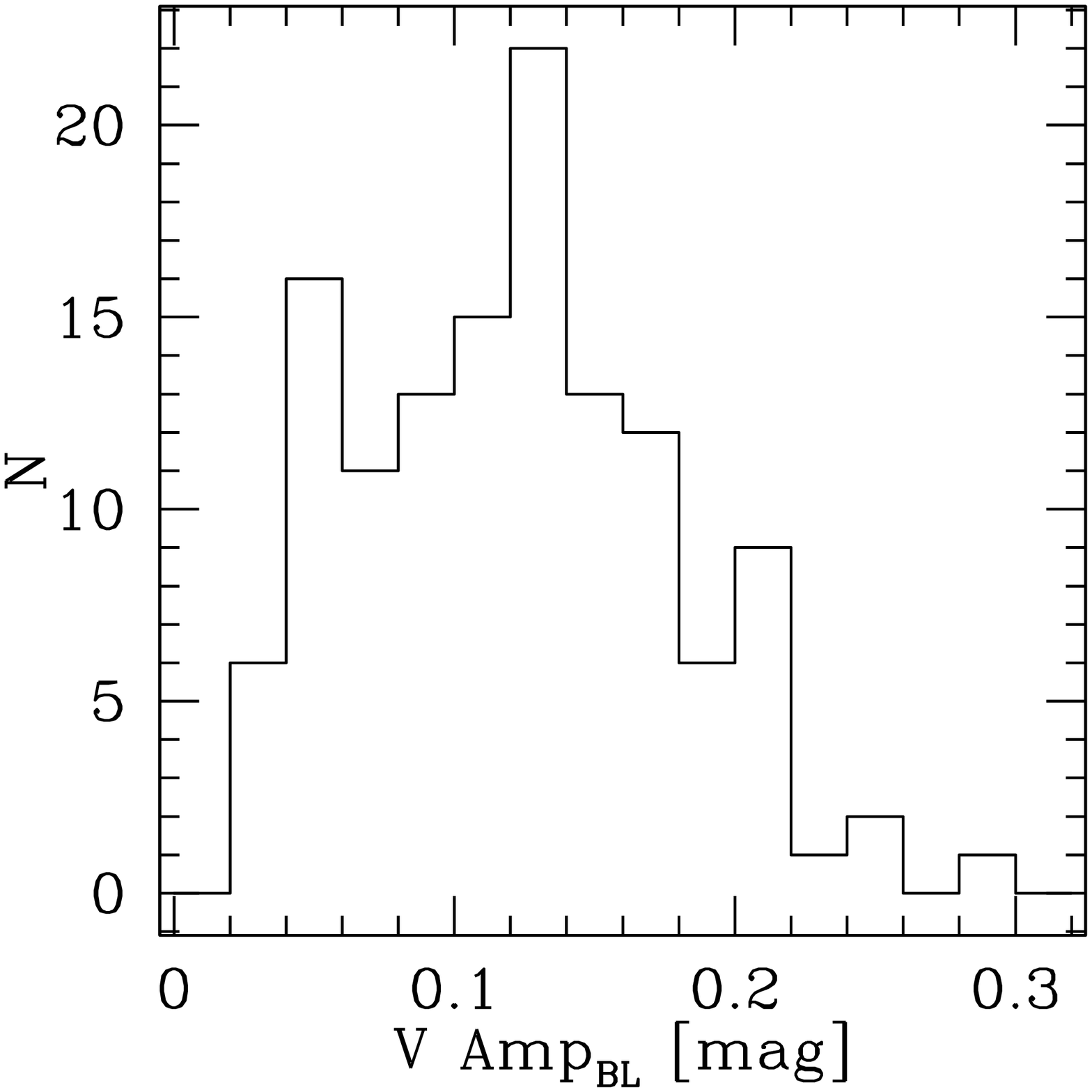} \\
\end{tabular}
\end{center}
\caption{The distribution of $V$ magnitudes of Blazhko stars (left), Blazhko periods (middle)
and Blazhko $V$ amplitudes (right) in the RR Lyrae sample. \label{bl.per.mag}}
\end{figure*}

We see that the number of Blazhko stars, namely 122 objects, is very low; we found the Blazhko effect in 5.1\% of RRab stars and 6.5\% of RRc stars, which is much lower 
than the number determined by Mizerski (2003) for the Galactic Bulge (25\% RRab and 10\% RRc)
and Soszynski et al. (2003) for the LMC (15\% RRab and 6\% RRc). 
The result for the SMC by Soszynski et al. (2002) of 10\% RRab and 10\% RRc is somewhat closer 
to our result, but the discrepancy is still visible.
The photometric precision of the data is a few percent, which suggests there is serious
incompleteness for the Blazhko effect at low amplitudes.
Figure \ref{bl.per.mag} (left panel) shows the distribution of mean $V$ magnitudes for Blazhko stars,
which is somewhat shifted towards the bright end relative to all RR Lyrae stars (Figure \ref{mag.amp}), suggesting that detection fails for faint stars, for which data scatter is higher. 

We also compared our results with previous BL searches in the sky surveys of the Galaxy.
In the recent paper on a similar data quality Northern Sky Variability Survey (Wils, Lloyd
and Bernhard 2006) only 34 out of 785 (4.3\%) stars exhibit the Blazhko effect,
although the time span of the data was just a year. In a Blazhko search in ASAS data
done by Wils and S{\'o}dor (2005), only 43 objects were identified ($\sim$ 4\%), though the search
did not include the latest part of ACVS with objects having declinations $\delta>0$ 
and inspected only RRab variables.

While the result in this paper is an improvement compared to numbers given by other Galactic field
investigations, the percentage of Blazhko stars is significantly lower than for other systems. 
What is even more striking, is that we found slightly more Blazhko object among RRc variables than 
among RRab which is opposite to the usual findings. We think that the result for the RRc group is real 
(the percentage of Blazhko is similar to the one in previous papers, that is 5-10\%), but for some 
reason we are missing Blazhko objects among RRab stars. It is possible that the RRabs with strong 
Blazhko effect have such scattered light curves that they are classified as different type variables by the
automatic ASAS classification.

\begin{figure}
\vspace{-0.5cm}
\begin{center}
\includegraphics[scale=.40]{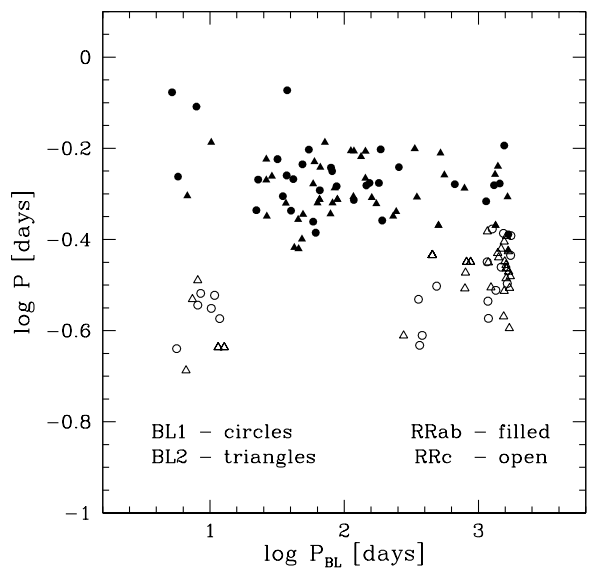}
\end{center}
\caption{Main pulsation period vs. Blazhko period.  The Blazhko periods of RRc stars exhibit a strongly bimodal distribution. \label{bl_per.vs.per}}
\end{figure}

Figure \ref{bl_per.vs.per} plots the Blazhko period $P_{BL}$ versus main pulsation period $P$. 
It is interesting that there is a significant difference in behaviour of RRab (filled symbols) 
and RRc (open symbols) groups. RRab stars have a relatively uniform distribution in the logarithm
of the Blazhko period, 
whereas RRc stars occupy either a short $P_{BL}$ region around 10 days, or a long 
$P_{BL}$ region, starting around 300 days with a condensation roughly around 1500 days. That is, 
we did not observe Blazhko changes with periods above $\sim$~20~days and below $\sim$~300~days 
in RRc pulsators. The unusual concentration around $P_{BL} 1500$~days, which is often close to 
the time span of the data, is adding to the peak near $log(P_{BL})=3$ in Figure \ref{bl.per.mag}
(middle panel). 
This concentration may be a false signal due to overall bias and those object might in fact be just exhibiting 
long period changes (PC). But there are objects having $P_{BL}$ values of 300-500~days, which is 
4-5 times shorter than the data span, thus the gap cannot be easily explained as a pile-up of PCs
misclassified as BLs.

Jurcsik et al. (2005) found a correlation between the pulsation period and the Blazhko period, such 
that RR Lyrae with periods $P < 0.4$~d can have $P_{BL}$ of order of days, while those with longer pulsation
periods such as $P > 0.6$~d always exhibit modulation with $P_{BL} > 20$~d. Our Figure \ref{bl_per.vs.per} 
does not support this statement. They also suggested that there is a ``continuous transition'' between 
BL1 and BL2 type modulations. 
While we observe differences in Blazhko behaviour between RRab and RRc groups, which can be due to 
physical differences in pulsation mechanisms in these groups, 
we note that BL1 (circles) and BL2 (triangles) occur in roughly constant proportion throughout Figure \ref{bl_per.vs.per}.
This observation may suggest that both BL1 and BL2 effects have the same origin, it is just that in case of BL1 
we do not observe the second equidistant peak present in BL2 category.

\begin{figure}
\begin{center}
\includegraphics[scale=.90]{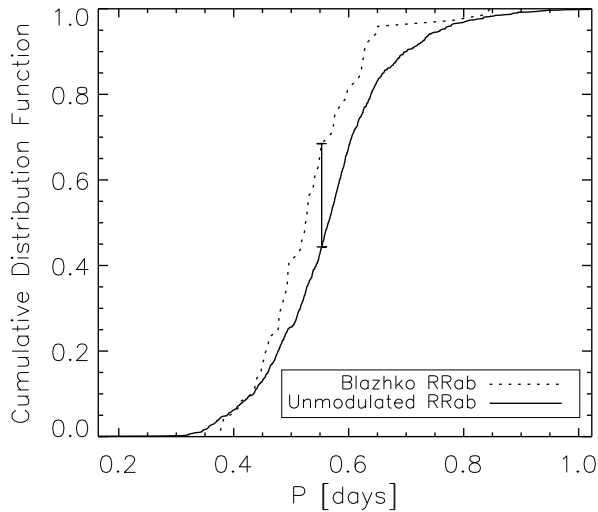}
\end{center}
\caption{The cumulative distribution function of the main pulsation period for RRab Blazhko stars and those
RRab that are not modulated.  The maximum difference in the curves is marked with a bar and is highly significant.  The lower incidence of the Blazhko effect for stars with long pulsation periods could give fundamental insight to the physical cause of the Blazhko effect, but we suspect that they exist but dropped out of our sample due to misclassification (see text). \label{per.cdf}}
\end{figure}

Figure \ref{bl_per.vs.per} also reveals that the number of BL stars falls rapidly for pulsation
periods $P > 0.65$ d ($log(P) > -0.2$). We compared the period distributions for RRab stars exhibiting the 
Blazhko effect with those that do not. Figure \ref{per.cdf} is a plot of the cumulative distribution 
function of the main pulsation period for Blazhko RRab stars and for RRab stars whose light curves are not modulated.
The maximum difference is 0.241466 which occurs at $P=0.552945$ d and has 
a false alarm probability of $4.8 \times 10^{-4}$ according to the Kolmogorov-Smirnov test, as computed with the KSTWO 
function from Press et al. (1989). 
This observation could be an important constraint for models of the Blazhko effect, but it may also indicate misclassification of RRab stars that are modulated.  That is, light curves
of BL RRab are not as well-defined as their non-Blazhko counterparts and so have different Fourier parameters, 
on which the classification of Pojma\'nski (2002) was based. The risk of misclassification is especially strong
for long period BL RRab, which may appear as $\delta$ Cephei first overtone pulsators if their fourth harmonic is reduced relative to the second harmonic (see Figure 6 of Pojma\'nski, 2002).
If this explanation is true, it would help account for the low percentage of RRab BL stars with respect to other surveys, since there are some Blazhko stars that were not found because we accepted the automatic classification.

On the other hand, this observation is more likely a confirmation of a previously recognized effect, that the mean main pulsation period of Blazhko RRab stars is lower than that of the unmodulated RRab stars (Preston 1964, Smith 1981, Nemec 1985, Gloria 1990).  This effect has started to be considered in models of the Blazhko effect (Stothers 2006), and the current study is the most statistically significant detection of the effect to date.

\begin{figure}
\vspace{-0.5cm}
\begin{center}
\includegraphics[scale=.40]{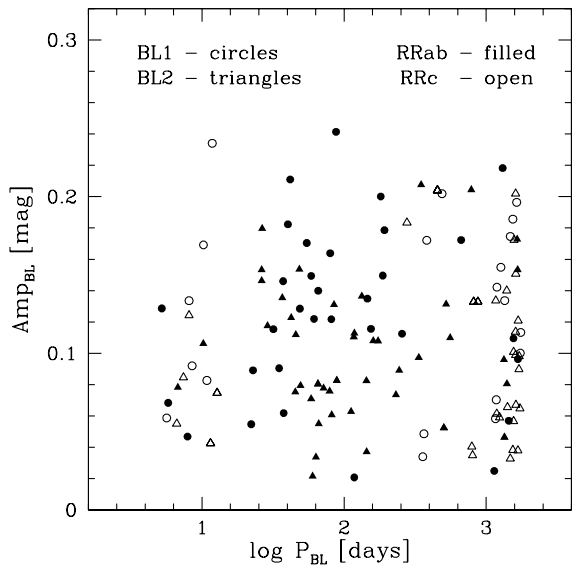}
\end{center}
\caption{Blazhko amplitude versus Blazhko period.
\label{bl.per.amp}}
\end{figure}

We also checked if there is any dependence of Blazhko amplitude on pulsation period. We do not see 
any significant trend; the distribution seems to be uniform. The right panel in Figure \ref{bl.per.mag} 
shows the distribution of modulation amplitudes.
Similarly, Blazhko amplitude versus modulation period (Figure \ref{bl.per.amp}) does not present 
any relation. Both long and short Blazhko periods can have an amplitude as small as 0.02 mag and as 
big as 0.2 mag.

The detailed results of our searches are listed in Table \ref{table.bl1} for BL1 
stars, Table \ref{table.bl2} for BL2 and Table \ref{table.pc} for PC objects.

Table \ref{table.bl1} contains magnitude, periods and amplitudes for BL1 stars. The main 
pulsation period is $P=f_{0}^{-1}$, where $f_{0}$ is the main pulsation frequency,
and Blazhko period is defined as $P_{BL}=(f_{1}-f_{0})^{-1}$, 
where $f_{1}$ is the frequency component caused by the Blazhko effect. Thus the negative value of 
$P_{BL}$ means that the additional frequency is smaller than the main pulsation frequency. 
This occurs in 10 out of 20 cases of RRc type and only 7 out of 32 of RRab, consistent with
previous statistics (eg. Moskalik and Poretti, 2003, Mizerski 2003)

\begin{table*}
\begin{scriptsize}
\caption{Objects exhibiting BL1 effect. $A_{0}$ is the peak-to-peak amplitude of variation as provided by ACVS 
and $A_{1}$ is the peak-to-peak amplitude of Blazhko modulation. See section \ref{sec:OeBe} for details.}
\begin{center}
\begin{tabular}{cccrccccccc}
\label{table.bl1}
 ASAS ID      &   $V$  &   $P$     &$P_{BL}$& $A_{0}$ & $A_{1}$& $A_{1}/A_{0}$ & $A_{noise}$ &Type & Other ID & \\  
              & [mag]  &  [days]   & [days] &  [mag]  &  [mag] &                 &             &     &          & \\
\hline
001141-0144.9 & 11.76 & 0.5297456 & 154.53 & 0.97 & 0.12 & 0.12 & 0.03 & RRAB & RY~Psc & \\
003338-1529.2 & 11.09 & 0.5737309 & -255.50 & 0.64 & 0.11 & 0.18 & 0.04 & RRAB & RX~Cet & * \\
003706-4317.7 & 13.29 & 0.6275343 & 187.12 & 1.31 & 0.15 & 0.11 & 0.04 & RRAB & - & \\
011831-6755.1 & 11.42 & 0.4057948 & 1748.86 & 0.49 & 0.11 & 0.23 & 0.04 & RRC & AM~Tuc& * \\
013140-4957.3 & 12.12 & 0.4604329 & 40.17 & 1.10 & 0.18 & 0.17 & 0.02 & RRAB & NSV00539 & * \\
020752-2651.9 &  9.48 & 0.4954335 & 34.88 & 1.13 & 0.09 & 0.08 & 0.02 & RRAB & SS~For & \\
022637-4119.7 & 10.08 & 0.2941932 & 357.94 & 0.13 & 0.03 & 0.26 & 0.01 & RRC: & - & *\\
025021-6415.7 & 12.80 & 0.5724975 & -79.81 & 1.29 & 0.16 & 0.13 & 0.04 & RRAB/DCEP-FO & RV~Hor & *\\
031113-2629.0 & 11.43 & 0.5973130 & 31.81 & 1.02 & 0.12 & 0.11 & 0.02 & RRAB & RX~For & \\
033108+0713.4 & 10.64 & 0.5281963 & -1442.79 & 0.18 & 0.06 & 0.32 & 0.02 & RRAB/EC/ESD & - & \\
053022-3234.8 & 11.77 & 0.2331114 & 364.60 & 0.20 & 0.05 & 0.24 & 0.01 & RRC/EC & - & \\
054230-1622.9 & 11.85 & 0.5389135 & 22.84 & 0.86 & 0.09 & 0.10 & 0.03 & RRAB & - & *\\
055322-5417.9 & 12.82 & 0.2452638 & 381.58 & 0.56 & 0.17 & 0.31 & 0.05 & RRC/EC & - & *\\
061401-6128.4 &  9.31 & 0.4857445 & -117.90 & 0.46 & 0.02 & 0.04 & 0.01 & RRAB & ST~Pic& * \\
062326+0005.8 & 11.96 & 0.5501317 & 37.24 & 0.63 & 0.15 & 0.23 & 0.04 & RRAB & - & \\
064615-4319.2 & 13.05 & 0.3186021 & 1639.61 & 0.71 & 0.20 & 0.28 & 0.07 & RRC & - & \\
070854+1919.7 &  8.69 & 0.7789192 & 7.92 & 0.10 & 0.05 & 0.45 & 0.02 & RRAB & - & *\\
071549-4405.3 & 13.04 & 0.3144609 & -488.40 & 0.63 & 0.20 & 0.32 & 0.05 & RRC & - & \\
081933-2358.2 & 10.45 & 0.2856671 & -8.10 & 0.29 & 0.13 & 0.47 & 0.06 & RRC/EC & - & *\\
085254-0300.3 & 12.42 & 0.2669022 & -11.80 & 0.51 & 0.23 & 0.46 & 0.09 & RRC/EC/ESD & - & *\\
090900-0410.4 & 10.68 & 0.3032613 & -8.52 & 0.41 & 0.09 & 0.22 & 0.04 & RRC & - & *\\
093731-1816.2 & 13.05 & 0.5209175 & 87.73 & 1.23 & 0.24 & 0.20 & 0.06 & RRAB & - & \\
101200+1921.9 & 10.52 & 0.4826394 & -1141.03 & 0.85 & 0.02 & 0.03 & 0.13 & RRAB & -& \\
123812-4422.5 & 12.69 & 0.5235490 & 1307.70 & 1.28 & 0.22 & 0.17 & 0.06 & RRAB: & - & \\
124805-0820.8 & 11.98 & 0.5105287 & 65.69 & 0.88 & 0.14 & 0.16 & 0.04 & RRAB & - & \\
135740-1202.3 & 12.18 & 0.2671226 & 1189.91 & 0.44 & 0.14 & 0.32 & 0.03 & RRC & - & \\
135813-4215.1 & 12.43 & 0.5231816 & -146.01 & 0.84 & 0.13 & 0.16 & 0.04 & RRAB & - & \\
141025-2244.8 & 12.50 & 0.6398808 & 1556.66 & 0.82 & 0.11 & 0.13 & 0.03 & RRAB & - & \\
144154-0324.7 & 11.40 & 0.2293674 & -5.65 & 0.32 & 0.06 & 0.18 & 0.01 & RRC/EC & - & * \\
145315-1435.9 & 12.43 & 0.5400738 & 41.77 & 0.94 & 0.21 & 0.23 & 0.05 & RRAB & - & \\
155553-4041.7 & 11.54 & 0.5819812 & 48.82 & 0.72 & 0.13 & 0.18 & 0.05 & RRAB & NSV07330 & * \\
170223-2422.0 & 11.34 & 0.4613693 & 22.18 & 0.40 & 0.05 & 0.14 & 0.01 & RRAB & - & \\
172721-5305.9 & 12.30 & 0.4354330 & 58.66 & 1.25 & 0.15 & 0.12 & 0.03 & RRAB & - & \\
180023-7026.5 & 12.08 & 0.3556146 & 1162.79 & 0.46 & 0.06 & 0.13 & 0.02 & RRC & - & \\
181215-5206.9 & 12.59 & 0.8375462 & -5.22 & 0.67 & 0.13 & 0.19 & 0.01 & RRAB & - & \\
185719-6321.4 & 12.28 & 0.4120017 & 61.39 & 0.96 & 0.12 & 0.13 & 0.03 & RRAB & - & \\
194502+2434.2 & 11.72 & 0.8458661 & 37.56 & 0.30 & 0.06 & 0.20 & 0.01 & RRAB/EC/ESD & - & \\
200431-5352.3 & 10.95 & 0.3002402 & 10.82 & 0.31 & 0.08 & 0.26 & 0.04 & RRC & - & *\\
200556-0830.9 & 12.75 & 0.4381966 & 192.20 & 1.12 & 0.18 & 0.16 & 0.04 & RRAB & KM~Aql & \\
202746-2850.5 & 11.93 & 0.4084525 & 1674.48 & 0.52 & 0.10 & 0.19 & 0.03 & RRAB/EC/ESD & - & \\
203420-2508.9 & 11.58 & 0.5262389 & 666.44 & 0.90 & 0.17 & 0.19 & 0.05 & RRAB & - & \\
203749-5735.5 & 12.25 & 0.4199162 & -1270.33 & 0.43 & 0.15 & 0.36 & 0.07 & RRC/EC & - & *\\
210741-5844.2 & 13.22 & 0.3462376 & 1479.95 & 0.68 & 0.17 & 0.26 & 0.06 & RRC/EC & - & *\\
211839+0612.3 & 11.06 & 0.2914601 & -1176.75 & 0.48 & 0.07 & 0.15 & 0.01 & RRC & - & *\\
212034+1837.2 & 11.50 & 0.5624065 & 81.30 & 0.76 & 0.12 & 0.16 & 0.03 & RRAB & - & \\
212331-3025.0 & 12.29 & 0.3674420 & 1739.74 & 0.47 & 0.10 & 0.21 & 0.03 & RRC/EC & - & \\
213826-3945.0 & 12.97 & 0.4107031 & -1540.12 & 0.62 & 0.19 & 0.30 & 0.09 & RRC & - & *\\
221556-2522.6 & 11.30 & 0.5467383 & 5.78 & 0.78 & 0.07 & 0.09 & 0.01 & RRAB & - & \\
225248-2442.2 & 12.78 & 0.5295565 & 181.20 & 1.19 & 0.20 & 0.17 & 0.05 & RRAB & - & \\
230659-4354.6 & 12.66 & 0.2811062 & -10.24 & 0.39 & 0.17 & 0.43 & 0.05 & RRC/EC/ESD & BO~Gru &* \\
231209-1855.4 & 12.57 & 0.3079943 & -1349.89 & 0.47 & 0.13 & 0.29 & 0.05 & RRC/EC & - & *\\
232031-1447.9 & 12.46 & 0.6269552 & 54.52 & 0.59 & 0.17 & 0.29 & 0.04 & RRAB & - & *\\
\hline
\end{tabular}
\end{center}
\end{scriptsize}
\end{table*}

Tables \ref{table.bl2} contains 
magnitude, periods and amplitudes for BL2 stars. The main pulsation period is also 
$P=f_{0}^{-1}$, where $f_{0}$ is the main pulsation frequency.
Now the Blazhko period is defined as $P_{BL}=(f_{-,+}-f_{0})^{-1}$ where $f_{-,+}$ 
means we choose between Blazhko frequencies ($f_{-}$ and $f_{+}$, with $f_{-} < f_{0} < f_{+}$)
the one with higher amplitude. The negative value of $P_{BL}$ means that the Blazhko peak of higher 
amplitude has a frequency lower than the main pulsation. This corresponds to the ratio of 
$A_{-}/A_{+}>1$. We see that this case occurs in 13 out of 29 objects of RRc type 
and 19 out of 41 of RRab. 

\begin{table*}
\begin{scriptsize}
\caption{Objects exhibiting BL2 effect. $A_{0}$ is the peak-to-peak amplitude of variation as provided by ACVS 
and $A_{-}$ and $A_{+}$ are the peak-to-peak amplitudes of Blazhko modulation corresponding to the modulation 
frequency on the left and right side of the main pulsation frequency, respectively. 
See section \ref{sec:OeBe} for details.}
\begin{center}
\begin{tabular}{cccrccccccc}
\label{table.bl2}
   ASAS ID    &   $V$    &    $P$      &$P_{BL}$  &$A_{0}$  & $A_{-}$ & $A_{+}$ &$A_{-}/A_{+}$&$A_{noise}$& Type & Other ID \\  
              & [mag]  &  [days]   & [days]   & [mag]   & [mag]   &  [mag]  &               &           &      &  \\
\hline
003514-0415.0 & 12.76 & 0.3445751 & -1616.29 & 0.60 & 0.14 & 0.15 & 0.96 & 0.03 & RRC/EC & - \\
012848-1127.2 & 12.59 & 0.5166664 & -84.99 & 1.21 & 0.13 & 0.12 & 1.06 & 0.01 & RRAB/DCEP-FO & - \\
020728-5752.2 & 10.88 & 0.3750261 & 1673.36 & 0.44 & 0.04 & 0.04 & 1.11 & 0.00 & RRC & NSV00728 \\
021515-1048.0 & 10.52 & 0.6234139 & 112.05 & 0.64 & 0.10 & 0.06 & 1.55 & 0.01 & RRAB & RV~Cet \\
030534-3116.1 & 12.53 & 0.4964538 & -6.77 & 1.47 & 0.08 & 0.06 & 1.24 & 0.02 & RRAB & - \\
031408-3446.4 & 11.41 & 0.3124235 & -1241.77 & 0.52 & 0.06 & 0.04 & 1.34 & 0.01 & RRC & - \\
032438-2334.7 & 12.09 & 0.6296339 & 335.29 & 0.94 & 0.12 & 0.10 & 1.26 & 0.01 & RRAB & - \\
032520-6503.3 & 11.24 & 0.4920082 & -160.64 & 0.88 & 0.11 & 0.10 & 1.03 & 0.04 & RRAB/DCEP-FO: & X~Ret \\
050747-3351.9 & 12.05 & 0.4873552 & 65.41 & 1.01 & 0.08 & 0.10 & 0.77 & 0.03 & RRAB & SU~Col \\
050747-3351.9 & 12.05 & 0.4873552 & 88.98 & 1.01 & 0.08 & 0.08 & 0.98 & 0.03 & RRAB & SU~Col \\
050838-5602.9 & 12.10 & 0.5160761 & -786.91 & 1.08 & 0.20 & 0.05 & 3.79 & 0.03 & RRAB & NSV~1856 \\
051508-4137.7 & 10.45 & 0.4788368 & 81.95 & 0.84 & 0.15 & 0.06 & 2.44 & 0.02 & RRAB & RY~Col \\
052402-2247.4 & 13.40 & 0.6498517 & -10.23 & 1.19 & 0.10 & 0.11 & 0.94 & 0.03 & RRAB & - \\
052840-5316.2 & 13.38 & 0.3678062 & -453.43 & 0.67 & 0.20 & 0.17 & 1.21 & 0.05 & RRC & - \\
053628-3837.0 & 12.64 & 0.3714727 & -1394.31 & 0.57 & 0.09 & 0.14 & 0.61 & 0.01 & RRC & - \\
054843-1627.0 & 12.88 & 0.3767273 & 1663.06 & 0.88 & 0.15 & 0.17 & 0.93 & 0.03 & RRAB/RRC: & - \\
070001-3732.5 & 11.85 & 0.4941843 & 116.96 & 0.89 & 0.13 & 0.11 & 1.21 & 0.02 & RRAB & NSV03331 \\
080249-5913.5 & 11.75 & 0.3541891 & -1185.26 & 0.28 & 0.06 & 0.06 & 0.90 & 0.01 & RRC/EC/ESD & - \\
080318-2530.1 & 10.86 & 0.2697723 & -1542.02 & 0.43 & 0.04 & 0.03 & 1.30 & 0.01 & RRC & NSV03882 \\
085448-8317.0 & 12.27 & 0.4778598 & -36.79 & 1.30 & 0.14 & 0.10 & 1.40 & 0.03 & RRAB & NSV04350 \\
091349-0919.1 & 10.71 & 0.5372290 & 26.30 & 0.90 & 0.15 & 0.16 & 0.97 & 0.03 & RRAB/DCEP-FO & SZ~Hya \\
094438-4552.6 & 11.58 & 0.5735076 & 66.35 & 0.95 & 0.05 & 0.06 & 0.98 & 0.01 & RRAB & CD~Vel \\
103203-3010.6 & 11.56 & 0.3304459 & 1731.60 & 0.47 & 0.06 & 0.06 & 1.00 & 0.01 & RRC & NSV04885 \\
103246-3423.1 & 13.39 & 0.2448541 & -277.05 & 0.78 & 0.12 & 0.18 & 0.65 & 0.04 & RRC & - \\
105303-4954.4 & 10.76 & 0.5274139 & 58.68 & 0.85 & 0.08 & 0.07 & 1.17 & 0.02 & RRAB & AF~Vel \\
110522-2641.0 & 11.68 & 0.2944559 & -7.40 & 0.38 & 0.08 & 0.07 & 1.14 & 0.02 & RRC & - \\
112027-4338.8 & 11.16 & 0.3795948 & 1480.38 & 0.28 & 0.03 & 0.04 & 0.92 & 0.03 & RRC/EC/ESD & - \\
114555-5922.7 & 11.16 & 0.4531949 & -79.01 & 1.19 & 0.07 & 0.08 & 0.90 & 0.02 & RRAB & BI~Cen \\
120447-2740.7 &  9.76 & 0.6503243 & 71.79 & 0.62 & 0.13 & 0.08 & 1.61 & 0.03 & RRAB & IK~Hya \\
120942-3457.4 & 13.20 & 0.3436271 & 1611.60 & 0.66 & 0.15 & 0.10 & 1.47 & 0.03 & RRC & EL~Hya \\
121206-2612.8 & 12.65 & 0.3987747 & 48.46 & 1.31 & 0.20 & 0.15 & 1.27 & 0.04 & RRAB & - \\
123030-2602.9 &  9.81 & 0.4785475 & 63.29 & 0.94 & 0.05 & 0.03 & 1.41 & 0.01 & RRAB & SV~Hya \\
132922-0553.0 & 11.24 & 0.5763497 & 1398.99 & 0.87 & 0.14 & 0.08 & 1.76 & 0.04 & RRAB & - \\
140324-3624.4 & 10.71 & 0.4939666 & -1650.71 & 0.98 & 0.15 & 0.17 & 0.86 & 0.04 & RRAB & V0674~Cen \\
141345-2254.7 & 12.05 & 0.4479434 & 26.50 & 1.00 & 0.30 & 0.18 & 1.65 & 0.04 & RRAB & - \\
150924-4319.6 & 12.47 & 0.3821508 & 42.49 & 1.15 & 0.26 & 0.12 & 2.08 & 0.04 & RRAB & FU~Lup \\
151849-1000.0 & 12.03 & 0.3364272 & 802.95 & 0.50 & 0.17 & 0.03 & 4.84 & 0.08 & RRC & - \\
153517-2420.2 & 11.24 & 0.3067770 & 1560.06 & 0.54 & 0.06 & 0.06 & 1.10 & 0.02 & RRC & CG~Lib \\
153830-6906.4 & 12.26 & 0.6224747 & -118.05 & 0.89 & 0.11 & 0.08 & 1.36 & 0.02 & RRAB & - \\
155552-2148.6 & 11.38 & 0.2541338 & 1699.52 & 0.46 & 0.10 & 0.09 & 1.12 & 0.02 & RRC & - \\
160204+1728.8 & 10.72 & 0.2308114 & 11.49 & 0.44 & 0.04 & 0.05 & 0.82 & 0.04 & RRC & LS~Her \\
160204+1728.8 & 10.72 & 0.2308114 & 12.76 & 0.44 & 0.07 & 0.10 & 0.74 & 0.04 & RRC & LS~Her \\
162158+0244.5 & 12.47 & 0.3238044 & -8.11 & 0.52 & 0.12 & 0.12 & 0.99 & 0.03 & RRC/EC & - \\
162811+0304.3 & 13.09 & 0.5970104 & -26.28 & 1.65 & 0.14 & 0.15 & 0.92 & 0.03 & RRAB & - \\
163225-8354.2 &  8.99 & 0.5425804 & -143.73 & 0.85 & 0.08 & 0.08 & 0.99 & 0.04 & RRAB & UV~Oct \\
172932-5548.3 & 12.89 & 0.3273052 & -1610.05 & 0.68 & 0.20 & 0.17 & 1.19 & 0.02 & RRC & EZ~Ara \\
174048-3132.6 & 10.60 & 0.4272967 & -504.03 & 0.85 & 0.05 & 0.04 & 1.17 & 0.02 & RRAB & V0494~Sco \\
174202-4633.7 & 10.72 & 0.3115788 & -1706.78 & 0.49 & 0.10 & 0.10 & 1.01 & 0.03 & RRC & - \\
175911-4926.0 & 10.11 & 0.4518593 & -49.37 & 0.94 & 0.08 & 0.08 & 0.97 & 0.02 & RRAB & S~Ara \\
183441-6527.0 & 11.88 & 0.4769536 & -173.70 & 1.17 & 0.11 & 0.09 & 1.22 & 0.02 & RRAB & BH~Pav \\
184758-3744.4 & 10.26 & 0.5893445 & -59.96 & 0.59 & 0.02 & 0.02 & 1.36 & 0.01 & RRAB & V0413~CrA \\
192824-1852.4 & 12.57 & 0.3563567 & 1572.33 & 0.61 & 0.17 & 0.19 & 0.93 & 0.05 & RRC & - \\
193538-7409.9 & 12.51 & 0.3499993 & 1608.49 & 0.60 & 0.11 & 0.12 & 0.97 & 0.02 & RRC/EC & - \\
195142-6244.1 & 11.86 & 0.5514395 & 557.17 & 1.14 & 0.12 & 0.11 & 1.10 & 0.02 & RRAB & FO~Pav \\
195927-3400.1 & 11.88 & 0.3797209 & 45.69 & 0.77 & 0.14 & 0.11 & 1.26 & 0.04 & RRAB & - \\
200910-4149.5 & 12.61 & 0.4401943 & 45.39 & 1.15 & 0.09 & 0.08 & 1.16 & 0.02 & RRAB & V2239~Sgr \\
202044-4107.1 & 10.93 & 0.5529452 & 1331.74 & 0.86 & 0.12 & 0.10 & 1.27 & 0.02 & RRAB & V1645~Sgr \\
203145-2158.7 & 11.25 & 0.3107152 & 792.83 & 0.39 & 0.04 & 0.04 & 1.11 & 0.01 & RRC/EC & - \\
204440-2402.7 & 12.75 & 0.2053330 & -6.64 & 0.36 & 0.06 & 0.04 & 1.26 & 0.01 & RRC/DSCT/EC & - \\
210129-1513.8 & 10.47 & 0.4477465 & 231.66 & 1.04 & 0.17 & 0.07 & 2.35 & 0.02 & RRAB & RV~Cap \\
212433-5712.1 & 12.95 & 0.6051401 & -133.38 & 1.14 & 0.14 & 0.09 & 1.54 & 0.02 & RRAB & - \\
214101+0109.6 & 12.43 & 0.6156709 & -522.58 & 0.57 & 0.13 & 0.10 & 1.36 & 0.02 & RRAB & - \\
214719-8739.1 & 12.34 & 0.4580005 & 244.20 & 1.25 & 0.09 & 0.09 & 1.01 & 0.04 & RRAB & RS~Oct \\
215336-8246.8 & 11.34 & 0.6218493 & 144.12 & 0.98 & 0.04 & 0.05 & 0.80 & 0.01 & RRAB & SS~Oct \\
220254-2131.5 & 10.73 & 0.3637073 & 1413.43 & 0.46 & 0.07 & 0.07 & 1.04 & 0.01 & RRC & BV~Aqr \\
222539-0756.5 & 10.66 & 0.4052637 & 1618.65 & 0.37 & 0.07 & 0.07 & 1.10 & 0.01 & RRC & GP~Aqr \\
225131-3006.2 & 13.10 & 0.3384769 & -1681.80 & 0.62 & 0.12 & 0.12 & 0.99 & 0.03 & RRC/EC & - \\
225323+0846.1 & 12.60 & 0.4930493 & -348.58 & 1.51 & 0.18 & 0.21 & 0.88 & 0.04 & RRAB & - \\
225518-2317.6 & 12.99 & 0.3935794 & 1557.88 & 0.62 & 0.13 & 0.10 & 1.29 & 0.01 & RRC/EC & - \\
233951-1644.4 & 12.16 & 0.3553741 &  875.96 & 0.52 & 0.17 & 0.13 & 1.26 & 0.06 & RRC & - \\
\hline
\end{tabular}
\end{center}
\end{scriptsize}
\end{table*}

Finally, Table \ref{table.pc} contains magnitude, period, and amplitude information 
for PC stars. The designations are the same as in Table \ref{table.bl1}. 
For RRab, equal numbers of stars have the modulation frequency lying on left and right side 
of the main pulsation frequency, which suggests
that PC stars may be physically distinct from BL1, and not simply BL1s with periods longer 
than the dataset.

\begin{table*}
\begin{scriptsize}
\caption{Objects exhibiting long period changes (PC). See section \ref{sec:OeBe} for details.}
\begin{center}
\begin{tabular}{cccrcccccc}
\label{table.pc}
  ASAS ID     &   $V$  &   $P$     &$P_{BL}$ & $A_{0}$ & $A_{1}$ &$A_{1}/A_{0}$& $A_{noise}$& Type & Other ID \\  
              & [mag]  &  [days]   & [days]  &  [mag]  &  [mag]  &               &            &      &	 \\
\hline
010117-4556.6 & 11.70 & 0.3738269 &  2085.07 & 0.27 & 0.08 & 0.29 & 0.02 & RRC & - \\
011831-6755.1 & 11.42 & 0.4057948 & -1702.13 & 0.49 & 0.10 & 0.21 & 0.04 & RRC & AM~Tuc \\
032337+1540.0 &  9.66 & 0.2384770 & -1738.83 & 0.13 & 0.02 & 0.17 & 0.00 & RRC/EC/ESD & - \\
033108+0713.4 & 10.64 & 0.5281963 &  1116.69 & 0.18 & 0.04 & 0.24 & 0.02 & RRAB/EC/ESD & - \\
040500-4457.1 & 12.66 & 0.5668385 & -1533.98 & 0.78 & 0.05 & 0.06 & 0.01 & RRAB & - \\
045213-5620.8 & 13.59 & 0.3503107 &  2050.02 & 0.75 & 0.21 & 0.28 & 0.06 & RRC & - \\
052406-6925.2 & 13.18 & 0.5531259 &  2668.80 & 2.21 & 0.18 & 0.08 & 0.02 & RRAB & TY~Dor \\
052840-5316.2 & 13.38 & 0.3678062 &  2178.17 & 0.67 & 0.16 & 0.24 & 0.05 & RRC & - \\
062838-3848.5 & 12.53 & 0.4834677 &  2282.06 & 1.41 & 0.10 & 0.07 & 0.03 & RRAB & - \\
065307-5935.7 & 11.47 & 0.7370705 & -2123.14 & 0.85 & 0.05 & 0.06 & 0.02 & RRAB & IU~Car \\
085816-3022.1 & 10.60 & 0.5107257 & -2403.27 & 0.12 & 0.03 & 0.22 & 0.01 & RRAB/DCEP-FO/EC/ESD & - \\
104749-0308.8 & 13.04 & 0.3632554 &  2236.14 & 1.02 & 0.14 & 0.13 & 0.06 & RRAB/DCEP-FO/EC & - \\
112027-4338.8 & 11.16 & 0.3795948 & -2346.87 & 0.28 & 0.04 & 0.14 & 0.03 & RRC/EC/ESD & - \\
112715-2510.4 & 13.39 & 0.5768075 & -2457.61 & 1.66 & 0.15 & 0.09 & 0.06 & RRAB/DCEP-FO & - \\
113701-0600.4 & 12.57 & 0.4022837 &  1880.76 & 0.60 & 0.09 & 0.15 & 0.01 & RRC/EC & - \\
140324-3624.4 & 10.71 & 0.4939666 & -1650.71 & 0.98 & 0.17 & 0.18 & 0.04 & RRAB & V0674~Cen \\
151849-1000.0 & 12.03 & 0.3364272 & -2192.98 & 0.50 & 0.03 & 0.07 & 0.08 & RRC & - \\
151951-0950.0 & 13.38 & 0.3235636 & -8591.07 & 0.70 & 0.00 & 0.00 & 0.02 & RRC: & - \\
154504+1736.7 & 12.82 & 0.5736250 &  1309.24 & 1.27 & 0.09 & 0.07 & 0.07 & RRAB & - \\
161301-2406.9 & 12.41 & 0.6024391 &  2036.66 & 0.43 & 0.07 & 0.17 & 0.02 & RRAB: & - \\
195927-3400.1 & 11.88 & 0.3797209 & -2285.19 & 0.77 & 0.12 & 0.16 & 0.04 & RRAB & - \\
210741-5844.2 & 13.22 & 0.3462376 & -1839.25 & 0.68 & 0.17 & 0.25 & 0.06 & RRC/EC & - \\
212045-5649.2 & 12.38 & 0.3680388 &  2286.76 & 0.26 & 0.05 & 0.20 & 0.02 & RRC/EC & - \\
212756-6702.6 & 12.73 & 0.3449728 &  2410.80 & 0.43 & 0.09 & 0.20 & 0.03 & RRC & - \\
220454-6635.0 & 12.05 & 0.5640177 &  1722.65 & 0.88 & 0.14 & 0.15 & 0.05 & RRAB & NSV14009 \\
221039-5049.8 & 12.32 & 0.3306480 & -1720.28 & 0.70 & 0.15 & 0.21 & 0.05 & RRC & - \\
224131-0628.6 & 11.86 & 0.5744469 & -1771.17 & 0.73 & 0.15 & 0.21 & 0.02 & RRAB & HH~Aqr \\
225559-2709.9 & 12.71 & 0.3103066 & -2197.80 & 0.37 & 0.08 & 0.22 & 0.02 & RRC/EC & - \\
232342-4157.4 & 12.72 & 0.4424689 & -2144.08 & 0.33 & 0.07 & 0.21 & 0.01 & RRC/EC/ESD & - \\
\hline
\end{tabular}
\end{center}
\end{scriptsize}
\end{table*}

Two stars (one RRab and one RRc) have two pairs of equidistant 
frequencies in the prewhitened spectrum. Their properties are listed in Table \ref{table.4_bl2x2}. 
The RRab star, SU Col, is discussed in the next section.  For the RRc star, LS Her, these pairs of 
BL2 peaks are close to one another, probably indicating that the Blazhko effect has an unstable period.

\begin{table*}
\begin{scriptsize}
\caption{RR Lyrae Blazhko stars with two pairs of BL2s. See section \ref{sec:OeBe} 
for details.}
\begin{center}
\begin{tabular}{cccccccccc}
\label{table.4_bl2x2}
    ASAS ID   &   $V$  &  $A_{0}$ & $P$      &$P_{BL}$& $A_{1}$ &  $A_{2}$& $A_{-}/A_{+}$ & Type & Other ID \\  
              & [mag]  & [mag]   & [days]    &[days]&  [mag]  &  [mag]  &         &      &         \\
\hline
050747-3351.9 & 12.05 & 1.01 & 0.4873552 & 65.8 & 0.08 & 0.10 & 0.77 & RRAB &  SU Col \\
              &       &      &           & 89.3 & 0.08 & 0.08 & 0.98 &      &         \\
\hline
160204+1728.8 & 10.72 & 0.44 & 0.2308114 & 11.5 & 0.04 & 0.05 & 0.82 & RRC  &  LS Her \\
              &       &      &           & 12.8 & 0.07 & 0.10 & 0.74 &      &         \\
\hline
\end{tabular}
\end{center}
\end{scriptsize}
\end{table*}

\subsection{A star with a rich Blazhko spectrum}
\label{sec:3bl}

In the course of this analysis we identified one star (ASAS 050747-3351.9 = SU Col) that has two clearly 
distinct pairs of BL2 peaks.  In order to look for additional extra frequencies, we first folded the raw 
light curve and determined, based on the scatter, that a 6 harmonic model would be appropriate.  
After removing those, we ran the CLEAN (Roberts et al. 1987) algorithm, with a gain of 0.5, for 200 iterations.
The frequency grid, of spacing $0.125/T$, ran from $0$~day$^{-1}$ to $10$~day$^{-1}$.  We then sorted the peaks
and identified pairs of peaks which were equidistant across each of the harmonics.  Additional frequency 
components of the Blazhko periods were found surrounding higher harmonics, as well as a candidate for a third 
Blazhko period.  The parameters returned by CLEAN were refined with the Levenberg-Marquardt algorithm according 
to the model
\begin{equation}
X_m(t) = \mu - \sum_{i=7}^{15} \frac{A_i}{2} \cos[ 2 \pi f_i (t - T_i) ], \label{eqn.bl2x2}
\end{equation}
where the mean magnitude was fit to be $\mu = 12.5980$ and the resulting fits for the other parameters 
($f_i$, $A_i$, and $T_i$) are in Table \ref{table.bl2x2}.

\begin{figure}
\includegraphics[scale=.90]{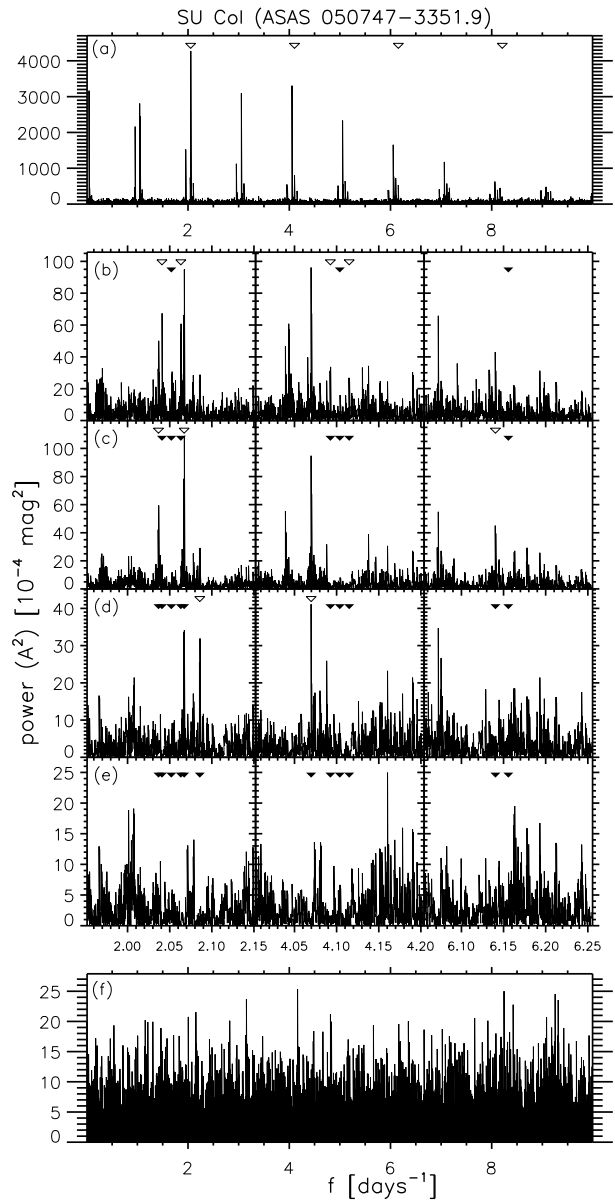} 
\caption{
An iterative subtraction of extra frequency components of SU Col illustrates the rich multiperiodic behaviour of this
Blazhko star.  (a) The Lomb-Scargle periodogram of the raw light curve.  (b) Periodogram after removing frequency
sinusoids at the main frequency and 5 higher harmonics.  The three panels are zoomed in on the region surrounding the
first three harmonics.  A number of Blazhko peaks are noticeable.  (c, d, e) Periodograms after removing the extra peaks
associated with $f_{B1}$, $f_{B2}$, and $f_{B3}$ respectively (see table 7).  (f) The final periodogram after subtracting
the 15 peaks listed in Table \ref{table.bl2x2}.  In each panel, open triangles identify the peaks that are most significant, and filled
triangles identify frequencies at which peaks have already been subtracted.
See section \ref{sec:3bl}.\label{fig:0507}}
\end{figure}

The identification and removal of these significant frequencies is illustrated in Figure \ref{fig:0507}.  
Panel (a) shows a Lomb-Scargle (Lomb 1976, Scargle 1982) periodogram of the raw light curve, with the first 
four harmonics marked with open triangles.  A six harmonic model was fitted by least squares, taking 
observational errors into account, and subtracted from these data.  Panels (b)-(e) show the periodogram 
of the residual light curves within $0.1$ day$^{-1}$ of each of the first three harmonics.  Marked with
filled triangles are frequencies corresponding to sinusoids that have already been subtracted, whereas 
marked with open triangles are frequencies identified as significant in the current panel.  
Panels (b), (c), and (d) correspond to the Blazhko frequencies of $f_{B1} = 0.0112$~day$^{-1}$, 
$f_{B2} = 0.0152$~day$^{-1}$, and $f_{B3} = 0.0339$~day$^{-1}$, matching Blazhko periods of 89.2 days, 
65.6 days, and 29.5 days, respectively.  Panels (e) and (f) are the periodogram of the final residuals, 
zoomed in on the first three harmonics and of the entire interval, respectively.  The two sets of BL2 
frequencies near the main peak---left hand side of panel (b)---is what led us to investigate this star 
in detail.  In some cases apparently significant peaks disappear upon the removal of another peak; this 
is a symptom of aliasing.  An example can be clearly seen when the peak at $4.04$ day$^{-1}$ in panel (c) 
disappears upon removal of the peak at $f_{11} = 2.037$ day$^{-1}$.  We used the CLEAN algorithm to avoid 
precisely this type of confusion.

Despite our efforts using CLEAN, the reality of the third Blazhko period is not conclusively established 
because there is a peak in the window function at $f' = 2.0027$ day$^{-1}$ which couples $f_{B2}$ to $f_{B3}$.
The following numerical relationships approximately hold:
\begin{eqnarray}
f_{15} - f' &\approx&f_{12} \label{eqn:blcov1}\\
f_{14} + f' &\approx& f_2 - f_{B2}.  \label{eqn:blcov2}
\end{eqnarray}
Equation \ref{eqn:blcov1} implies $f_{15}$ could be a nearly $2$ day$^{-1}$ alias of $f_{12}$.  In addition, 
we note that the fitted parameters of these sinusoids have a large covariance, about half the geometric mean 
of the individual parameters' variances.  Equation \ref{eqn:blcov2} gives the reason for this covariance, 
although our analysis did not identify a significant sinusoid at $f_2 - f_{B2}$; if it exists, it is 
apparently buried in noise.

The status of SU Col as an RR Lyrae variable was established by Gessner (1985).  Berdnikov and Turner (1996) 
presented an O-C diagram and refined the ephemeris of maximum light, giving:
\begin{equation}
\rmn{Max~HJD} = 2450363.870(2) + 0.48735842(11) \times E. \label{ephem}
\end{equation}

The uncertainties are in parentheses, indicating the precision of the final and final two digits for epoch and 
period, respectively.

The maximum with number $E = 5008$ is near the center of our dataset, and gives a time of maximum $C = 2452804.561$.  
Although the cadence of our dataset does not allow individual maxima to be found, the nearby maximum of the six 
harmonic model is at $O = 2452804.558 \pm 0.010$, matching the published ephemeris to within the error 
($O-C = 0.002 \pm 0.010$).  Our best-fitting main period of $0.487355$ d $\pm 2 \times 10^{-6}$ d, also marginally fits the ephemeris of equation \ref{ephem}.

The Hipparcos and Tycho Catalog (Perryman and ESA 1997) reported SU Col as an RR Lyrae as well, reporting the epoch 
of maximum as $O_H = 2448500.2120 \pm 0.0010$.  The uncertainty was only reported to order-of-magnitude, and is 
based on a three harmonic model; judging by the folded light curve it is probably at least a factor of a few too low.  
This corresponds to maximum number $E = -3825$, whose calculated value is $C_H = 2448500.2114$, giving 
$O-C = 0.0006 \pm 0.0022$.  Therefore equation \ref{ephem} satisfies three modern data sets over a span of about 
fifteen years, despite the complication that the Blazhko effect causes on the time of maximum light.

The status of SU Col as a Blazhko star has not been previously recognized.  Wils, P. and Sod{\'o}r (2005) analyzed 
the ASAS RR Lyrae stars for Blazhko behavior, although we suspect that SU Col was not analyzed because its folded 
light curve looks like it is affected by noise rather than the Blazhko effect, and visually inspecting the folded 
light curves was their first cut.  Kovacs (2005) also identified Blazhko stars in ASAS, however, he restricted 
his attention to stars whose metallicity was already measured.

Recently it was discovered that another star with multiple Blazhko periods, namely XZ Cyg, is changing 
its Blazhko period over time (LaCluyze et al. 2004). The authors discover that these changes are 
anticorrelated with observed changes in the primary period of this star.
We looked for a possible Blazhko period change for SU Col. In particular, we divided the light curve in two, but each section had the same period.  Also, an O-C diagram for this object does not show any long term primary period changes (although perturbations by the Blazhko effect are seen), so the multiple Blazhko periods of SU Col cannot be explained analogously to XZ Cyg behaviour.

Future observations with a high cadence, spanning a few hundred days, would be extremely valuable for understanding 
the multiply periodic nature of the Blazhko effect in this star.  Such a campaign has recently reported similar double Blazhko behavoir for UZ UMa (S{\'o}dor et al. 2006).  However, a campaign with telescopes distributed 
in longitude may be necessary to resolve all aliases.  SU Col is considerably brighter than the class of objects 
with two sets of BL2 peaks found by Collinge et al. (2006), so even time-resolved spectroscopy is not prohibitive.  
Discovering rare objects that are bright enough to follow up is one of the strengths of wide-field surveys like ASAS, and we urge follow-up observations of this object.

\begin{table}
\caption{Best-fitted parameters for the detected frequencies in the Blazhko star SU Col. The parameters listed below 
refer to equation \ref{eqn.bl2x2}.  Times of light maximum are listed as HJD-2450000.  See section \ref{sec:3bl}.}
\begin{center}
\begin{tabular}{c c c c}
\label{table.bl2x2}
$f$ (day$^{-1}$) & A (mag) & T (day) & note \\	
\hline
2.0518901 & 0.69 & 2804.6078 & $f_1$ \\
4.1037803 & 0.30 & 2804.5774 & $f_2 = 2 f_1$ \\
6.1556702 & 0.20 & 2804.5569 & $f_3 = 3 f_1$ \\
8.2075605 & 0.11 & 2804.5484 & $f_4 = 4 f_1$ \\
10.259451 & 0.08 & 2804.6462 & $f_5 = 5 f_1$ \\
12.311340 & 0.05 & 2804.6246 & $f_6 = 6 f_1$ \\
2.0407786 & 0.08 & 2804.6775 & $f_7 = f_1 - f_{B1}$ \\
2.0631858 & 0.09 & 2804.4963 & $f_8 = f_1 + f_{B1}$ \\
4.0926081 & 0.06 & 2804.6302 & $f_9 = f_2 - f_{B1}$ \\
4.1149220 & 0.06 & 2804.5249 & $f_{10} = f_2 + f_{B1}$ \\
2.0366963 & 0.09 & 2804.5442 & $f_{11} = f_1 - f_{B2}$ \\
2.0671779 & 0.09 & 2804.5708 & $f_{12} = f_1 + f_{B2}$ \\
6.1403548 & 0.06 & 2804.5370 & $f_{13} = f_3 - f_{B2}$ \\
2.0857132 & 0.05 & 2804.6723 & $f_{14} = f_1 + f_{B3}$ \\
4.0697479 & 0.06 & 2804.5647 & $f_{15} = f_2 - f_{B3}$ \\
\hline
\end{tabular}
\end{center}
\end{table}

\section{Double-mode RR Lyrae}
\label{sec:rrd}

RRd variables are objects pulsating in two radial modes: the fundamental mode and the first overtone. 
Characteristic is the period ratio of these modes, $P_{1}/P_{0} \approx 0.744$, where $P_{0}$ and $P_{1}$ 
correspond to the periods of fundamental mode and the first overtone respectively. So far, 24 such 
isolated objects have been detected in the field of our Galaxy, 17 of them are listed by Wils (2006, see 
references therein) and 7 by Bernhard \& Wils (2006), without including fainter objects of the Galactic Bulge 
(Mizerski, 2003; Moskalik \& Poretti, 2003; Pigulski et al., 2003), and the foreground stars of the dwarf galaxy
in Sagittarius (Cseresnjes, 2001); 13 of them were found in the ASAS data. 

Very useful in the study of RRd stars is a Petersen diagram (Petersen 1973), on which the fundamental mode 
period ($P_{0}$) is plotted against the ratio of the overtone and fundamental mode periods ($P_{1}/P_{0}$).
Several attempts were made to constrain theoretically the region occupied by RRd stars on a Petersen diagram
(Bono et al. 1996, Popielski et al. 2000, Bragaglia et al. 2001, Kovacs 2001). Most recently Szabo et al. (2004)
combined both the double mode pulsational and evolutional theories to construct such a theoretically permitted
region.
We searched a slightly bigger region ($0.40 \leq P_{0} \leq 0.65$ and $0.735 \leq P_{1}/P_{0} \leq 0.755$) 
for RRd variables without putting any additional constraints. The condition that a star
had to fulfill to classify as a RRd candidate was that the ratio of the two highest amplitude
frequencies falls within the desired period ratio range. This procedure provided us with 35 candidates.

\begin{figure}
\begin{center}
\begin{tabular}{c c}
\includegraphics[scale=.30]{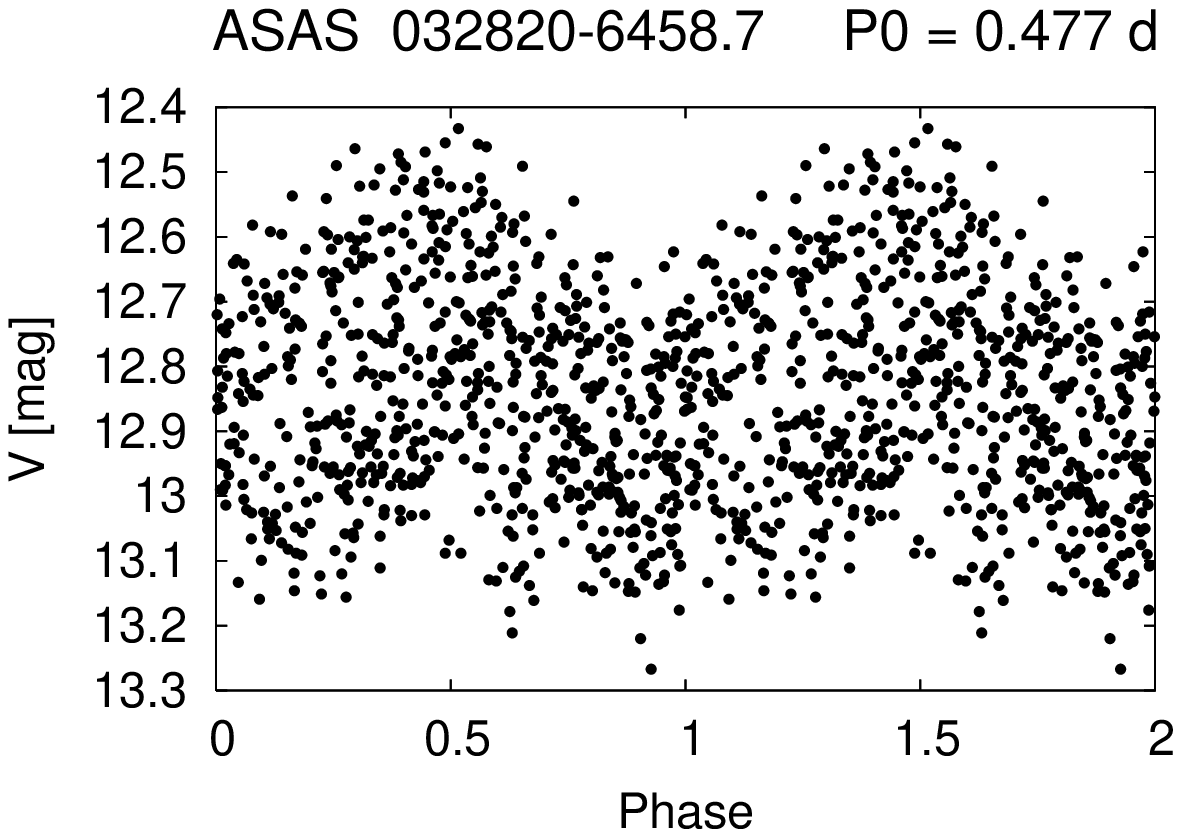} &
\includegraphics[scale=.30]{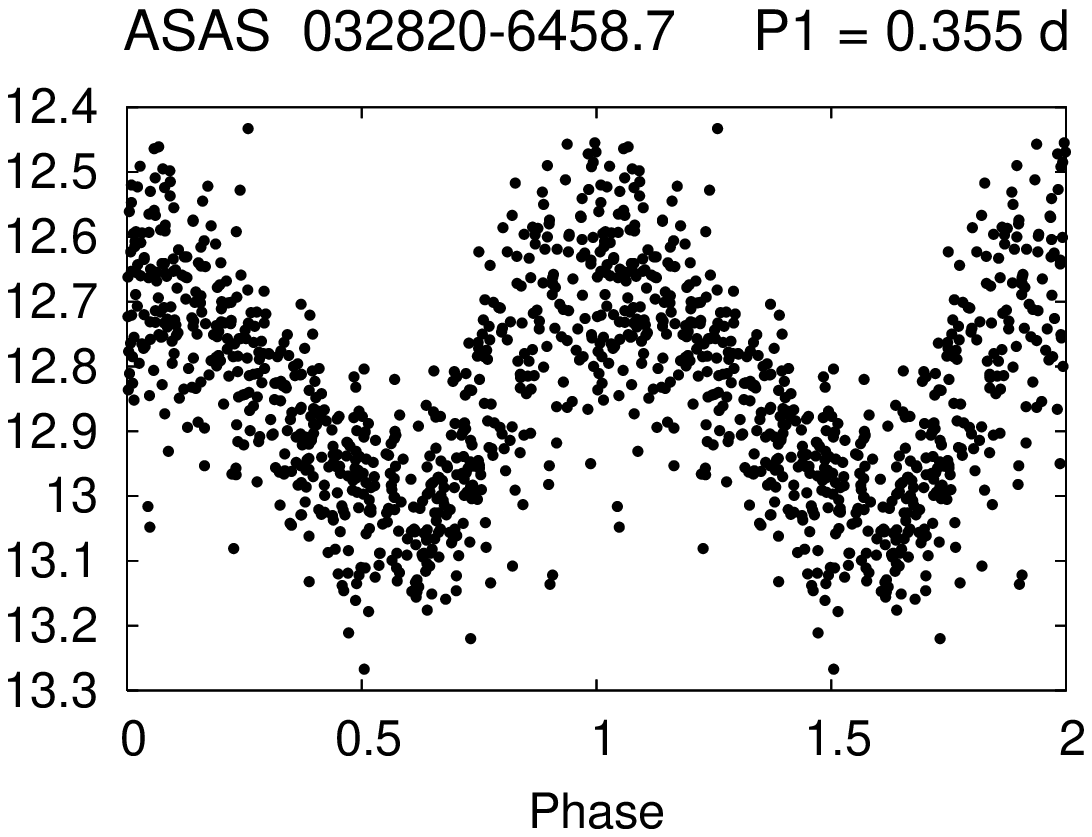} \\
\includegraphics[scale=.30]{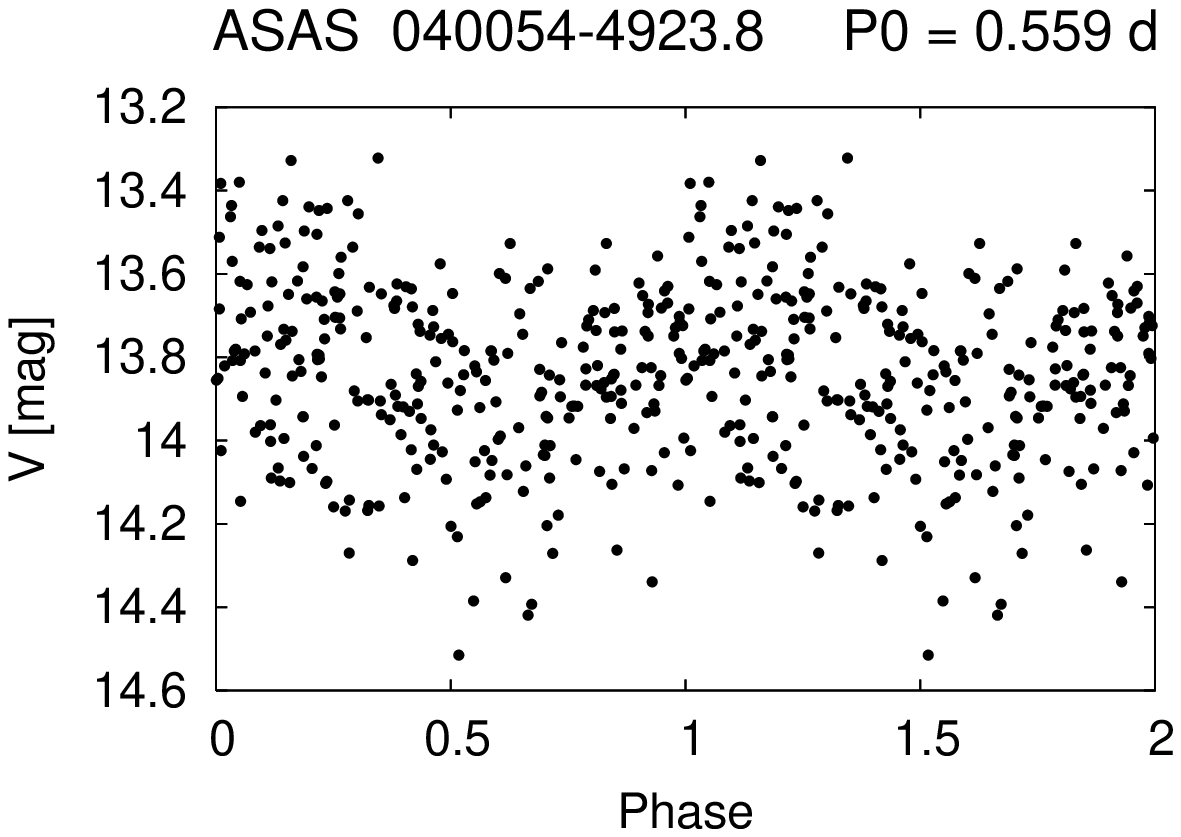} &
\includegraphics[scale=.30]{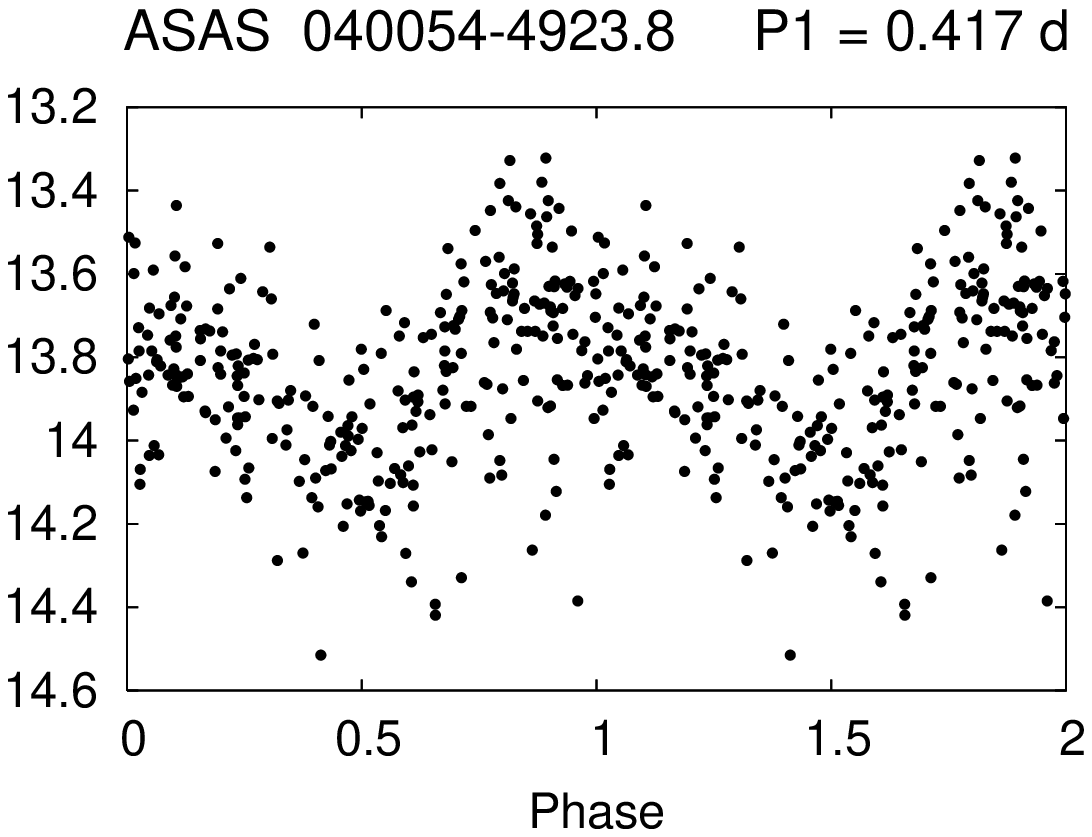} \\
\includegraphics[scale=.30]{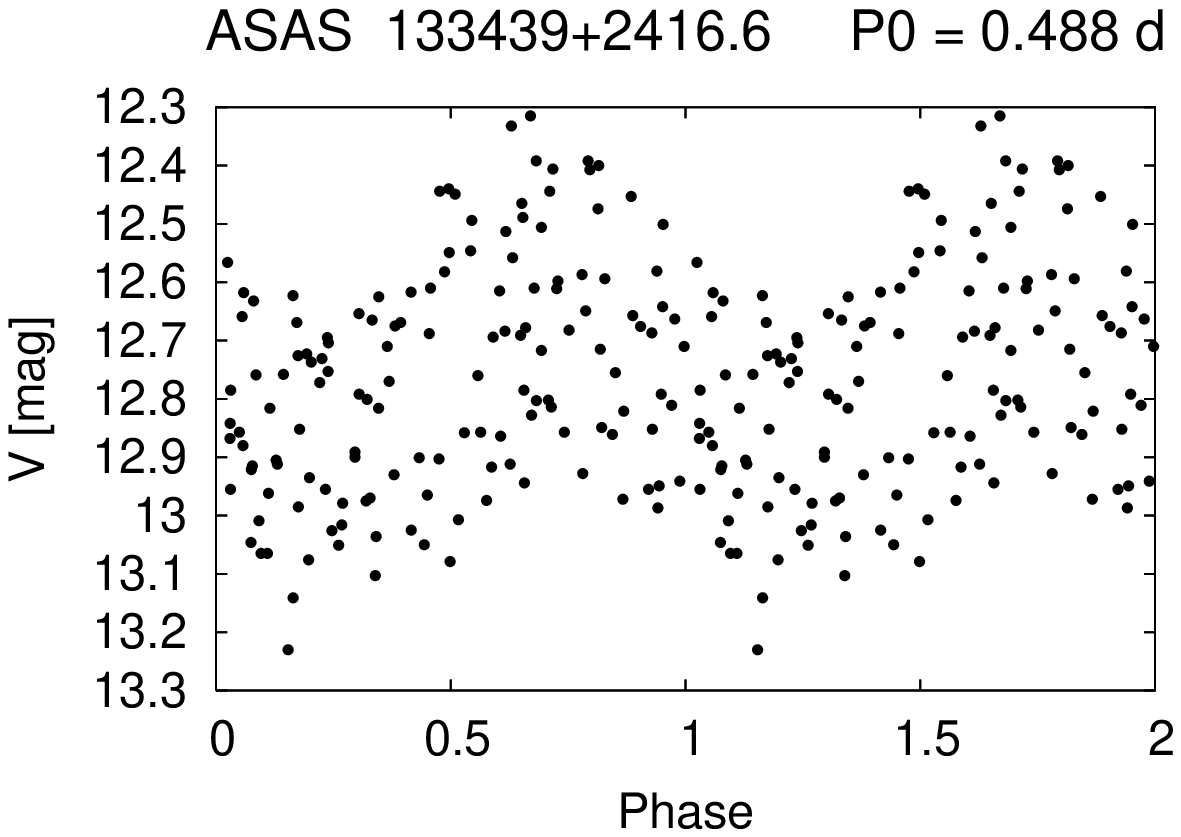} &
\includegraphics[scale=.30]{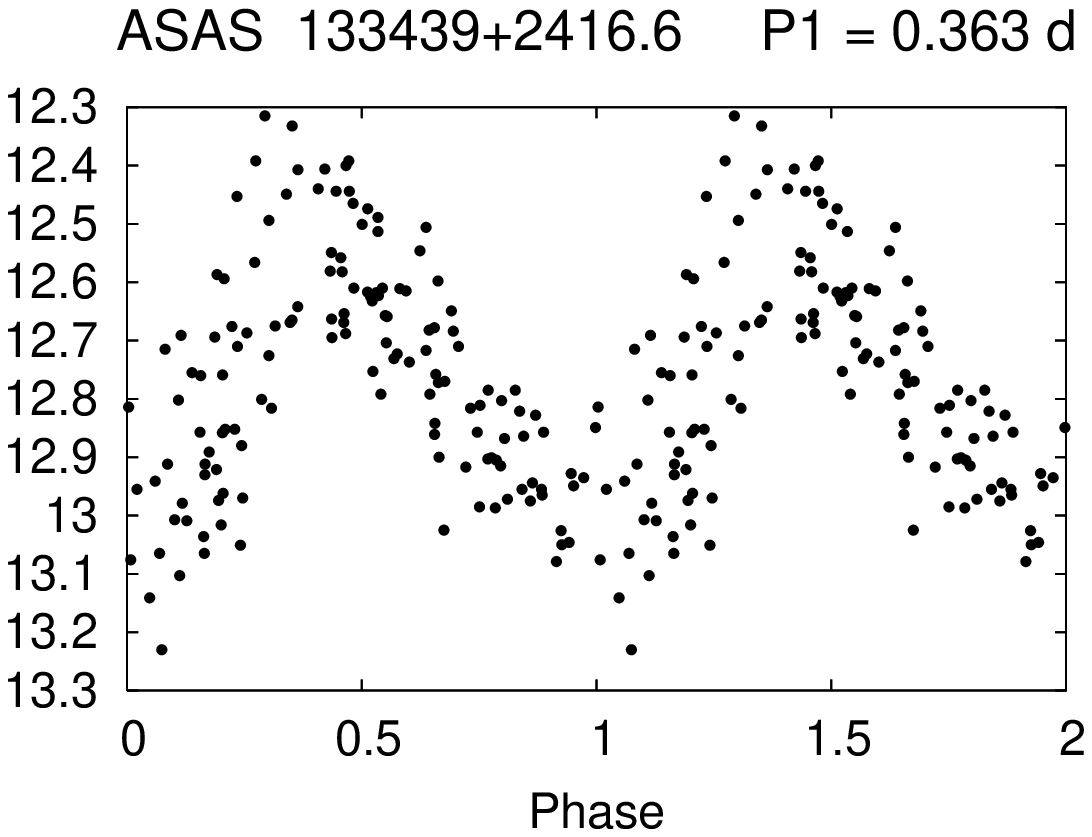} \\
\includegraphics[scale=.30]{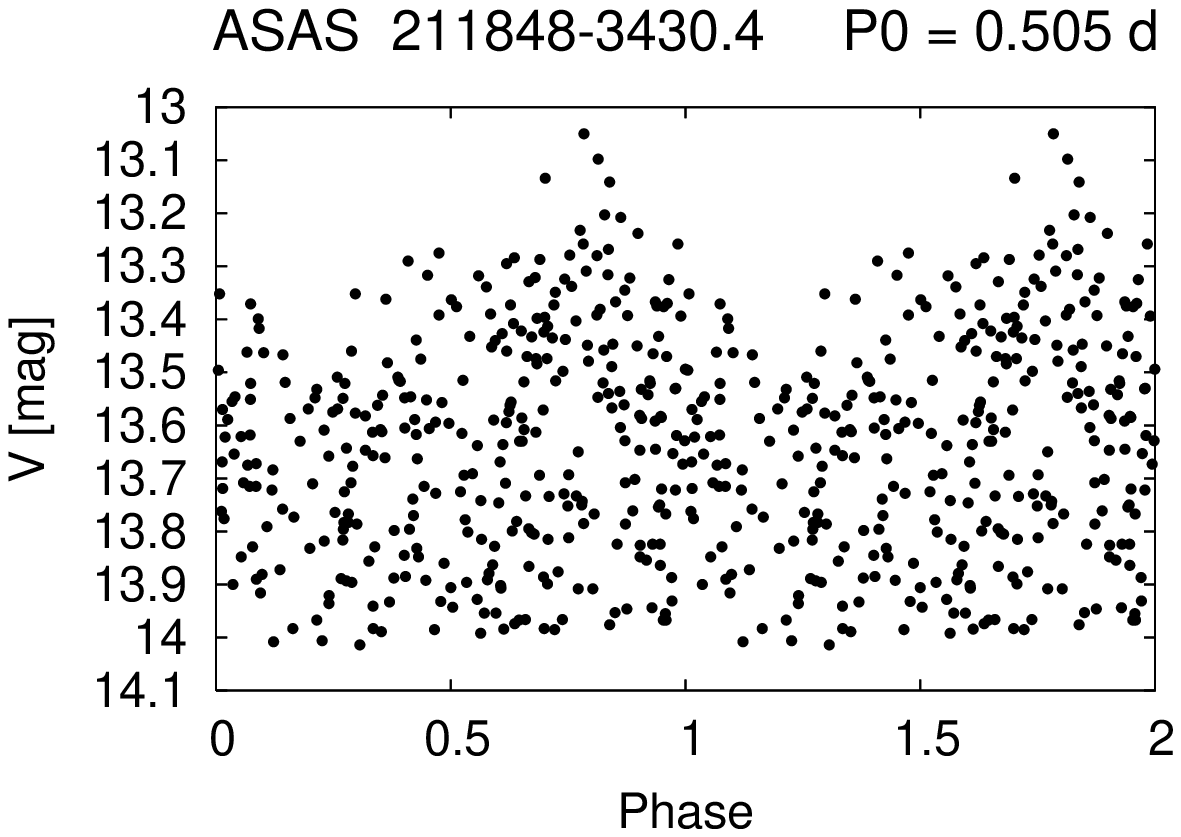} &
\includegraphics[scale=.30]{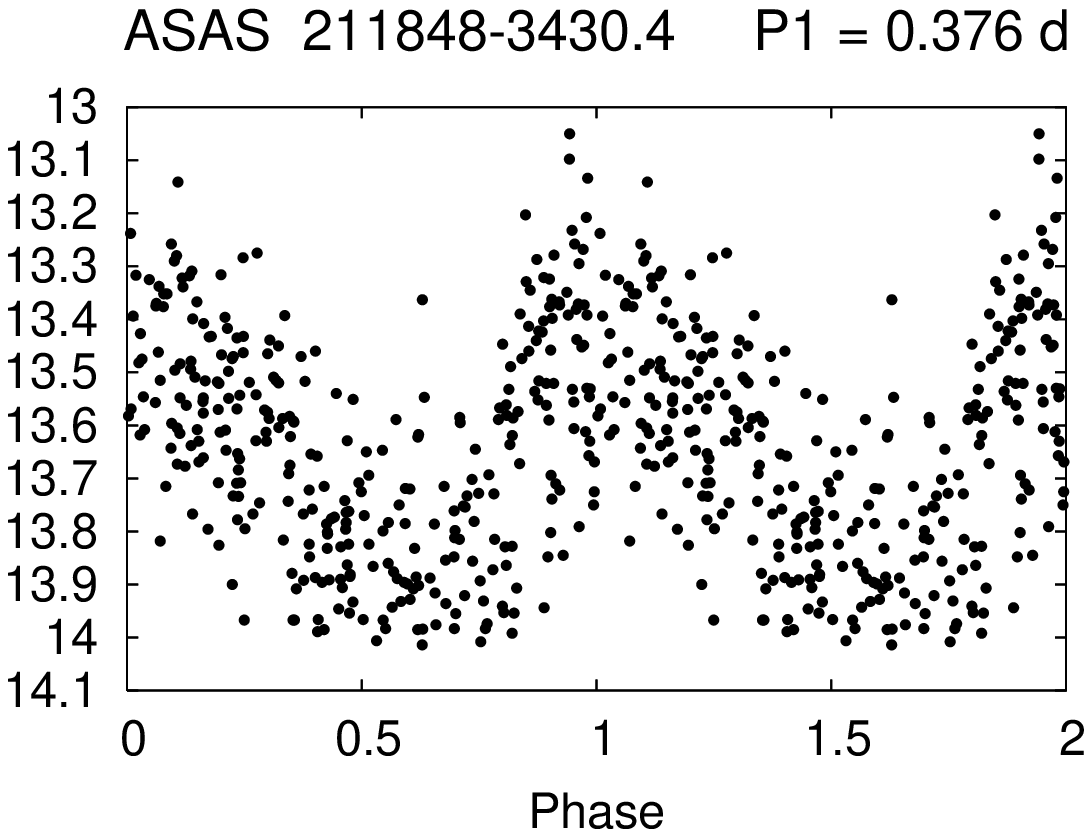} \\
\end{tabular}
\end{center}
\caption{Four newly discovered double-mode pulsators are represented in 4 rows. Each row contains a light curve
phased with the main pulsation period (left) and the first overtone period (right).
See section \ref{sec:rrd} for details. \label{double.mode}}
\end{figure}

We visually examined those and 16 turned out to be spurious detections. 
So in our search for multiperiodicity in ASAS data we detect 19 double mode RR Lyrae variables, 
among which 4 are newly identified RRd stars. This sums up to 17 RRd stars altogether discovered
in ACVS, out of 28 objects now known in the field of our Galaxy (on both southern and northern hemisphere), 
giving the ASAS project the majority of discoveries. The light curves of newly discovered objects, 
phased with both pulsation periods are given in Figure \ref{double.mode}.

Surprisingly, we did not detect 4 objects that were discovered as RRd before and should be present in the
ASAS database.
After a closer investigation of these objects we found that two of them, namely ASAS 030528-3058.7 
(Bernhard and Wils, 2006) and CU Com = ASAS 122447+2224.5 (Clementini et al., 2000) 
are phased in ACVS with one of the aliases of the true period, so the period ratio fell outside the searched region.
For the third one (Jerzykiewicz and Wenzel, 1977), namely AQ Leo = ASAS 112355+1019.0,
we did not find a significant peak among 50 highest that would correspond to the second pulsation period.
However, the light curve phased with this period has a modulation that appears significant, which apparently suggests that the CLEAN
algorithm failed in this case. The fourth case was an object found by Wils et al. (2006) in NSVS data in an 
overlap region, namely V458 Her = ASAS 170831+1831.3. We do not confirm any significant secondary frequency in this 
light curve with the longer dataset.

The incidence ratio, which we define as a number of double mode pulsators divided by the number of RRc
stars is RRd/RRc = 0.025. This value is small, comparing to the one determined by Alcock et al. (2000)
for the LMC, which is 0.134. The ratio for the Galactic bulge calculated by Mizerski (2003) is 0.007.

Parameters of these objects, such as their coordinates, magnitudes, pulsation periods and amplitudes,
are listed in Table \ref{table.rrd}. Also another star ID is given if applicable and a reference to the
discovery paper. For stars that are new discoveries, there is a ``*'' instead of a reference.
One of the newly found double mode stars, namely 032820-6458.7 (or SW Ret), is a secure detection based on its 
period ratio of 0.744417, but it is classified in GCVS as W UMa type eclipsing variable. However, due to slope
asymmetry, ACVS unambiguously classifies this object as an RRc star. 

It is known that the majority of RRd stars have a first pulsation mode with a lower amplitude than the first 
overtone (eg. Alcock et al. 2000).  The ASAS detections confirm this: 18 out of 19 objects have amplitude ratio 
$A_1/A_0 > 1$. 

\begin{figure}
\begin{center}
\includegraphics[scale=.35]{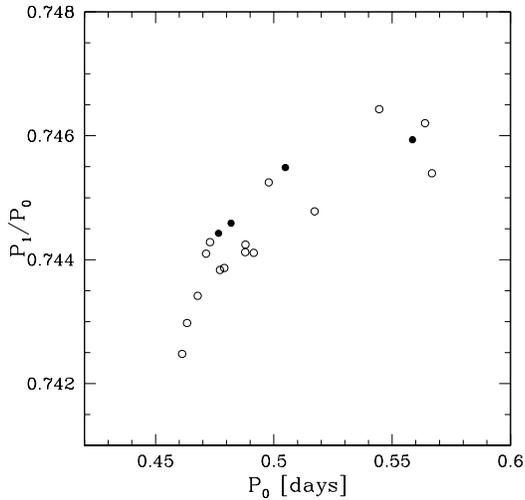}
\end{center}
\caption{A Petersen diagram for galactic field RRd variables discover during multiperiodicity search on ASAS data.
Known objects are plotted with open circles, while newly identified stars are drawn with filled circles.
 See section \ref{sec:rrd} for details. \label{petersen}}
\end{figure}

Figure \ref{petersen} shows a Petersen diagram for double-mode RR Lyrae in ASAS data. All stars follow
the sequence observed up to date by many RR Lyrae (see Figure 13 in Clementini et al. 2004).
The new discoveries on our diagram are plotted with solid circles and known objects with open circles. 

\begin{table*}
\caption{Parameters for double-mode RR Lyrae stars (RRd) found in ASAS data.  See section \ref{sec:rrd}.}
\begin{scriptsize}
\begin{tabular}{c c c c c c l c c c}
\label{table.rrd}
ASAS ID       &  $V$   & $P_{0}$   & $A_{0}$ & $P_{1}$   & $A_{1}$ &$P_{1}/P_{0}$&$A_{1}/A_{0}$& Other ID & Reference\\
(RA DEC)      & [mag]  & [days]    & [mag]   & [days]    & [mag]   &           &       &          &          \\
\hline
032820-6458.7 & 12.56 & 0.4766242 & 0.07 & 0.3548111 & 0.18 & 0.7444253 & 2.73 & SW Ret & * \\
040054-4923.8 & 13.32 & 0.5585883 & 0.10 & 0.4166705 & 0.16 & 0.7459348 & 1.61 & - & * \\
081610-6644.8 & 12.36 & 0.5172180 & 0.14 & 0.3852143 & 0.21 & 0.7447813 & 1.47 & GSC 8936-2145 & \cite{wo05}\\
084747-0339.1 & 10.53 & 0.5639063 & 0.08 & 0.4207890 & 0.20 & 0.7462037 & 2.59 & GSC 4868-0831 & \cite{wo05}\\
122509-2139.9 & 12.19 & 0.5445154 & 0.03 & 0.4064419 & 0.20 & 0.7464286 & 6.01 & - & \cite{bw06}\\\
133439+2416.6 & 12.44 & 0.4879017 & 0.12 & 0.3630583 & 0.23 & 0.7441218 & 1.85 & BS Com & \cite{brag03} \\
141539+0010.1 & 13.35 & 0.4819318 & 0.12 & 0.3588418 & 0.18 & 0.7445904 & 1.57 & - & * \\
151735-0105.3 & 10.96 & 0.4713494 & 0.15 & 0.3507304 & 0.21 & 0.7440985 & 1.41 & V0372 Ser & \cite{gm01} \\
173726+1122.4 & 12.76 & 0.4633236 & 0.17 & 0.3442391 & 0.17 & 0.7429777 & 1.02 & V2493 Oph & \cite{gm97} \\
183952-3200.9 & 11.81 & 0.4612583 & 0.20 & 0.3424752 & 0.18 & 0.7424802 & 0.89 & GSC 7411-1269 & \cite{wo05} \\
184035-5350.7 & 12.21 & 0.4790651 & 0.07 & 0.3563601 & 0.10 & 0.7438656 & 1.49 & - & \cite{bw06}\\
193933-6528.9 & 12.89 & 0.4915148 & 0.17 & 0.3657414 & 0.20 & 0.7441106 & 1.20 & GSC 9092-1397 & \cite{w06} \\
195612-5043.7 & 12.12 & 0.4678191 & 0.13 & 0.3477846 & 0.15 & 0.7434169 & 1.17 & GSC 8403-0647 & \cite{wo05} \\
210726+0110.3 & 13.06 & 0.4772373 & 0.17 & 0.3549859 & 0.22 & 0.7438352 & 1.27 & - & \cite{bw06}\\
211848-3430.4 & 12.84 & 0.5048597 & 0.10 & 0.3763662 & 0.22 & 0.7454867 & 2.12 & - & * \\
212721-1908.0 & 13.06 & 0.4730220 & 0.13 & 0.3520623 & 0.19 & 0.7442830 & 1.46 & NSV13710 & \cite{bw06}\\
213437-4907.5 & 12.00 & 0.4879954 & 0.14 & 0.3631872 & 0.19 & 0.7442432 & 1.38 & Z Gru & \cite{w06} \\
230449-3345.3 & 12.99 & 0.4978410 & 0.14 & 0.3710152 & 0.21 & 0.7452483 & 1.47 & - & \cite{bw06}\\
235622-5329.4 & 12.58 & 0.5668088 & 0.10 & 0.4224954 & 0.20 & 0.7453932 & 2.05 & NSV14764 & \cite{bw06}\\
\hline
\end{tabular}
\end{scriptsize}
\end{table*}

\section{Summary}
\label{sec:sum}

We performed a multiperiodicity search in the ASAS Galactic field RR Lyrae data, 
consisting of 1435 RRab and 756 RRc stars. 
It resulted in finding 160 objects, 122 of them exhibiting Blazhko effect (BL), 29 long period changes (PC), 
and 19 double-mode behaviour (RRd stars).

There are 73 BL stars inside the RRab group, and 49 in RRc, giving a total of 5.6\% BL in the whole sample, 
which is few times less than the percentage observed in Magellanic Clouds and Galactic Bulge, but similar to 
the number found in previous searches in Galactic field survey catalogs.
The analysis of the Blazhko sample reveals a significant difference between the behaviour of RRab and RRc 
groups. While RRab stars have a wide, unimodal distribution in the Blazhko period, RRc stars occupy 
either the short Blazhko period region around 10 days, or a long
$P_{BL}$ area, starting around 300 days, with a condensation roughly around 1500 days.  There is a lack of Blazhko stars among high period ($P > 0.65$ d) RRab stars, which may give insight into the physics of the Blazhko effect, but we suspect simply indicates the misclassification of RRab stars.
We do not confirm any correlation between the pulsation period and the Blazhko period.
The absence of visible differences between the BL1 and BL2 groups suggests that the same mechanism is responsible 
for the Blazhko effect in both cases.

During our study we discovered the Blazhko effect with multiple periods (BL2x2) in object ASAS
050747-3351.9 = SU Col. 
were identified with periodogram peaks near the first three harmonics of the main pulsation.

We also identified 19 double-mode RR Lyrae, 4 of them being new discoveries, one of these (SW Ret) being 
reclassified from an eclipsing contact binary to a RR Lyrae variable.

\section*{Acknowledgments}

We would like to thank G. Pojma{\'n}ski and B. Paczy{\'n}ski for being great advisors. We also thank T. Mizerski 
for helpful discussions. This research has made use of the SIMBAD database, operated at CDS, Strasbourg, France.
The work of DS was supported by the MNiSW grant N203 007 31/1328; DF was supported by NASA under award No NNG04H44G
to S. Tremaine.

\label{lastpage}


\begin{thebibliography}{}

\bibitem[()]{alco00}
Alcock, C., et al., 2000, ApJ, 542, 257

\bibitem[()]{al1co03}
Alcock, C., et al., 2003, ApJ, 598, 597

\bibitem[()]{bai1902}
Bailey, S.I., Pickering, E.C., 1902, Ann. Astron. Obs. Harvard Coll., 38, 1

\bibitem[()]{bt96}
Berdnikov, L.N. and Turner, D.G., 1996, IBVS, 4389, 1

\bibitem[Bernhard \& Wils 2006]{bw06}
Bernhard, K., Wils, P., 2006, IBVS 5698

\bibitem[Blazhko 1907]{b07}
Blazhko, S., 1907, Astron. Nachr., 175, 325

\bibitem[()]{bon96}
Bono, G., Incerpi, R., Marconi, M. 1996, ApJ, 467, 97

\bibitem[Bragaglia et al. 2003]{brag03}
Bragaglia et al., 2003, in: Annual Report, Osservatorio Astronomico di Bologna

\bibitem[()]{clem00}
Clementini G., di Tomaso S., di Fabrizio L., Bragaglia A., Merighi R., Tosi M., Caretta E., Gratton R.G., 
Ivans I.I., Kinard A., Marconi M., Smith H.A., Wilhelm R., Woodruff T., Sneden C., 2000, AJ, 120, 2054

\bibitem[()]{clem04}
Clementini, G., Corwin, T.M., Carney, B.W., Sumerel, A.N. 2004, AJ, 127, 938

\bibitem[()]{coll06}
Collinge, M., Sumi, T., and Fabrycky, D., 2006, ApJ, 651, 197

\bibitem[()]{cse01}
Cseresnjes P., 2001, A\&A, 375, 909

\bibitem[()]{duf06}
Duffau, S., Zinn, R., Vivas, A. K., Carraro, G., Mendez, R. A., Winnick, R., Gallart, C., 2006, ApJ, 636, 97

\bibitem[()]{edd17}
Eddington, A.S. 1917, Obs, 40, 290

\bibitem[Garcia-Melendo \& Clement 1997]{gm97}
Garcia-Melendo E., Clement M., 1997, AJ, 114, 1190

\bibitem[Garcia-Melendo et al. 2001]{gm01}
Garcia-Melendo E., Henden A.A., Gomez-Forrellad J.M., 2001, IBVS, 5167 

\bibitem[()]{gess85}
Gessner, H. 1985, MitVS, 10, 155

\bibitem[()]{glo90}
Gloria, K.A., 1990, PASP, 102, 338

\bibitem[()]{jerz}
Jerzykiewicz M., Wenzel W., 1977, AcA, 27, 35

\bibitem[()]{jur05}
Jurcsik, J., Szeidl, B., Nagy, A., S{\'o}dor, {\'A}., 2005, AcA, 55, 303

\bibitem[()]{kova01}
Kovacs, G. 2001, A\&A, 375, 469

\bibitem[()]{kova05}
Kovacs, G. 2005, A\&A, 438, 227

\bibitem[()]{lac04}
LaCluyze, A., Smith, H.A., Gill, E.-M., Hedden, A., Kinemuchi, K., Rosas, A.M., Pritzl, B.J., Sharpee, B.,
Wilkinson, C., Robinson, K.W., Baldwin, M.E., Samolyk, G., 2004, AJ, 127, 1653

\bibitem[()]{lomb76}
Lomb, N.R., 1976, Ap\&SS, 39, 447

\bibitem[()]{mize03}
Mizerski, T., 2003, AcA, 53, 307

\bibitem[()]{mos03}
Moskalik P., Poretti E., 2003, A\&A, 398, 213

\bibitem[()]{nemec85}
Nemec, J.M., 1985, AJ, 90, 204

\bibitem[()]{perr97}
Perryman, M.A.C., and ESA 1997, The HIPPARCOS and TYCHO catalogues (Noordwijk, Netherlands; ESA Publishing Division)

%

\bibitem[()]{pet73}
Petersen, J. O.,1973, A\&A, 27, 89

\bibitem[()]{pig03}
Pigulski A., Kolaczkowski Z., Kopacki G., 2003, Acta Astron., 53, 27

\bibitem[()]{Poj97}
Pojma\'nski, G.  1997, AcA, 47, 467

\bibitem[()]{Poj98}
Pojma\'nski, G.  1998, AcA, 48, 35

\bibitem[()]{Poj00}
Pojma\'nski, G.  2000, AcA, 50, 177

\bibitem[()]{Poj02}
Pojma\'nski, G.  2002, AcA, 52, 397

\bibitem[()]{Poj03}
Pojma\'nski, G.  2003, AcA, 53, 341

\bibitem[()]{Poj04}
Pojma\'nski, G., and Maciejewski, G.  2004, AcA, 54, 153

\bibitem[()]{Poj05}
Pojma\'nski, G., and Maciejewski, G.  2005, AcA, 55, 97

\bibitem[()]{PPS05}
Pojma\'nski, G., Pilecki, B., Szczygiel, D.  2005, AcA, 55, 275

\bibitem[()]{pop00}
Popielski, B.L., Dziembowski, W.A., Cassisi, S. 2000, AcA, 50, 491

\bibitem[()]{pres89}
Press, W.H., Teukolsky, S.A., Vetterling, W.T., \& Flannery, B.P. 1989, Numerical Recipes in C (2d ed.; Cambridge; Cambridge University Press)

\bibitem[()]{pres64}
Preston, G.W., 1964, ARA\&A, 2, 23

\bibitem[()]{robe87}
Roberts, D.H., Leh{\'a}r, J., \& Dreher, J.W., 1987, AJ, 93, 968

\bibitem[()]{scar82}
Scargle, J.D., 1982, ApJ, 263, 835

\bibitem[()]{shap14}
Shapley, H. 1914, ApJ, 40, 448

\bibitem[()]{shap18}
Shapley, H. 1918, ApJ, 48, 89

\bibitem[()]{smith81}
Smith, H. A. 1981, PASP, 93, 721

\bibitem[()]{smit95}
Smith, H.A., 1995, RR Lyrae Stars (Cambridge, UK: Cambridge University Press)

\bibitem[S{\'o}dor 2006]{s06}
S{\'o}dor, {\'A}., Vida, K., Jurcsik, J., V{\'a}radi, M., Szeidl, B., Hurta, Zs., D{\'e}k{\'a}ny, I., Posztob{\'a}nyi, K., Vityi, N., Szing, A., Kuti, A., Lakatos, J., Nagy, I., Dobos, V., 2006, IBVS, 5705

\bibitem[()]{sosz02}
Soszynski, I., Udalski, A., Szymanski, M., Kubiak, M., Pietrzynski, G., Wozniak, P., Zebrun, K., Szewczyk, O.,
Wyrzykowski, L., 2002, Acta Astron., 52, 369

\bibitem[()]{sosz03}
Soszynski, I., Udalski, A., Szymanski, M., Kubiak, M., Pietrzynski, G., Wozniak, P., Zebrun, K., Szewczyk, O., 
Wyrzykowski, L., 2003, Acta Astron., 53, 93

\bibitem[()]{stot06}
Stothers, R.B., 2006, ApJ, 652, 643

\bibitem[()]{sza04}
Szab{\'o}, R., Kolláth, Z., Buchler, J. R., 2004, A\&A, 425, 627

\bibitem[Wils 2006]{w06}
Wils, P., 2006, IBVS, 5685

\bibitem[Wils et al. 2006]{wlb06}
Wils, P., Lloyd, C., Bernhard, K. 2006, MNRAS, 368, 1757

\bibitem[WIls \& Otero 2005]{wo05}
Wils, P., Otero, S.A., 2005, IBVS, 5593, 1

\bibitem[Wils \& S{\'o}dor 2005]{ws05}
Wils, P. and S{\'o}dor, {\'A}., 2005, IBVS, 5655, 1

\end{thebibliography}
\end{document}